\documentclass[12pt]{iopart}

\usepackage{etoolbox}
\usepackage[switch]{lineno}

\usepackage[caption=false]{subfig}
\usepackage[hidelinks]{hyperref} 
\usepackage{breqn} 

\usepackage{standalone}
\usepackage{pgfplots}
\pgfplotsset{compat=newest}

\BeforeBeginEnvironment{dmath}{\begin{nolinenumbers}}
\AfterEndEnvironment{dmath}{\end{nolinenumbers}}

\begin{document}

\title[Capillary-Driven Kinetics]{Simulation of Capillary-Driven Kinetics with Multi-Phase-Field and Lattice-Boltzmann Method}
\author{Raphael Schiedung\(^1\), Marvin Tegeler\(^1\), Dmitry Medvedev\(^2\), Fathollah Varnik\(^1\)}
\address{\(^1\)Ruhr Universit\"at Bochum$,$ Interdisciplinary Center for Advanced Materials Simulation (ICAMS)$,$ Universit\"atsstr. 150$,$ 44801 Bochum$,$ Germany}
\address{\(^2\)Lavrentyev Institute of Hydrodynamics SB RAS$,$ Lavrentyev prosp.~15$,$ 630090$,$ Novosibirsk$,$ Russia}
\address{\(^2\)Novosibirsk State University$,$ Pirogova str.~2$,$ 630090$,$ Novosibirsk$,$ Russia}
\ead{raphael.schiedung@rub.de, fathollah.varnik@rub.de}
\begin{abstract}
    We propose a combined computational approach based on the multi-phase-field and the lattice Boltzmann method for the motion of solid particles under the action of capillary forces.
    The accuracy of the method is analyzed by comparison with the analytic solutions for the motion of two parallel plates of finite extension connected by a capillary bridge.
    The method is then used to investigate the dynamics of multiple spherical solid bodies connected via capillary bridges.
    The amount of liquid connecting the spheres is varied, and the influence of the resulting liquid-morphology on their dynamics is investigated.
    It is shown that the method is suitable for a study of liquid-phase sintering which includes both phase transformation and capillary driven rigid body motion.
\end{abstract}

\vspace{2pc}

\submitto{\MSMSE}

\section{Introduction}

During the sintering process, liquids can form intricate structures and bridges between multiple solid particles which by themselves can have a complex topology and can form large structures with other solid grains.
These capillary bridges between the solid particles lead to compaction of the sample.
Various theoretical approaches consider the force exerted by capillary bridges on solids.
An early analysis of capillary forces in liquid-phase sintering processes of spherical particles can be found in~\cite{heady_analysis_1970}.
The elementary capillary bridge between two identical spheres offers already a wide range of possible investigations such as its shape, the capillary force exerted by the bridge on a spherical particle, or the wetting angle and the amounts of liquid for which the bridge exists before its point of rupture~\cite{lian_theoretical_1993,rabinovich_capillary_2005,megias-alguacil_capillary_2009}.
A more complex scenario is considered in recent works~\cite{chen_liquid_2011,payam_capillary_2011} where also the effect of unequal sized spheres is investigated.

Villanueava et.\ al~\cite{villanueva_multicomponent_2009} proposed a combined approach of a multicomponent and multi-phase-field model with the Navier-Stokes equations for the simulation of liquid-phase sintering,
Also within the lattice Boltzmann methods~\cite{medvedev_simulating_2013,subhedar_modeling_2015}, models have been proposed to simulate the dynamics of solid particles, multiple fluids~\cite{stratford_lattice_2005,stratford_parallel_2008}, and more recently with the integration of liquid-gas interfaces with solid bodies~\cite{connington_interaction_2015}.
Lately, Sun and Sakai~\cite{sun_direct_2016} have performed an intensive numerical study on capillary bridges between two, three, and four spherical particles, where they used the so-called direct numerical simulation method.
In the spherical case of a capillary bridge between two spherical bodies, they also studied the motion of the bodies under the action of the capillary force.

However, we are not aware of any work addressing the full multiphysics problem of liquid phase sintering, simultaneously accounting for capillary forces, rigid body motion and phase transformation kinetics at solid-liquid and solid-solid interfaces.
Therefore, we present a combined approach of the so-called multi-phase-field method and the lattice Boltzmann method (Sec.~\ref{sec:model}).
We show the reliability of the model, by comparing the obtained results with analytic solutions for the force and motion of two finite parallel plates connected by a cylindrical liquid bridge.
The analytic solutions are introduced in Sec.~\ref{sec:theory}.
In addition to the comparison with the analytic solution, we show an investigation of the resolution dependence of the capillary force acting on the two plates, their motion due to the action of the capillary force and the wetting angle (see Sec.~\ref{sec:results_plate}).
Beyond these benchmark scenarios, we study the dynamics of two, three, and four spherical bodies under the action of the capillary force for various amounts of liquid fractions in Sec.~\ref{sec:results_dynamics}.
By doing this investigation, we show the differences in dynamics but also the similarities.
We summarize the results in the Sec.~\ref{sec:summary}.

\section{Theory}
\label{sec:theory}
As a test for our model, we consider the force of a single capillary bridge on two parallel plates and the resulting motion of these plates in Sec.~\ref{sec:theory_plate}.
An introductive example is the topology of a capillary bridge between two spherical bodies which depends on the amount of liquid and the wetting angle.
We discuss this dependency in Sec.~\ref{sec:theory_2_spheres}.

\subsection{Capillary Bridge Between Two Plates}
\label{sec:theory_plate}
In principle, a capillary bridge between two plates can form a variety of different surface shapes with a constant curvature which depends on the wetting angle, the distance between the plates, and the amount of liquid between the plates.
These shapes can be separated into two basic classes, convex and concave bridges.
In addition to this classification, analytic descriptions of these shapes can be found, such as a sphere, a cylinder, or a catenoid~\cite{kralchevsky_particles_2001}.

Here, we consider only simple examples to test our model, a planar capillary bridge with a \(90^\circ\) wetting angle and a cylindrical bridge with the same wetting angle.
For simplicity, the planar capillary bridge can be considered as a 2D problem.
Nevertheless, the planar capillary bridge example is simulated in a three-dimensional setup with periodic boundaries.

This approach not only offers the possibility to calculate the resulting force on the plates (see. Sec.~\ref{sec:theory_plate_force}) but also allows the integration of the equation of motion in order to obtain an approximate solution for the dynamics at early times as the plates start to move toward each other (see. Sec.~\ref{sec:theory_plate_motion}).

\subsubsection[Force]{Force on the Plates}
\label{sec:theory_plate_force}
\begin{figure}
    \subfloat[\label{fig:plate_setup_planar}]{\includegraphics{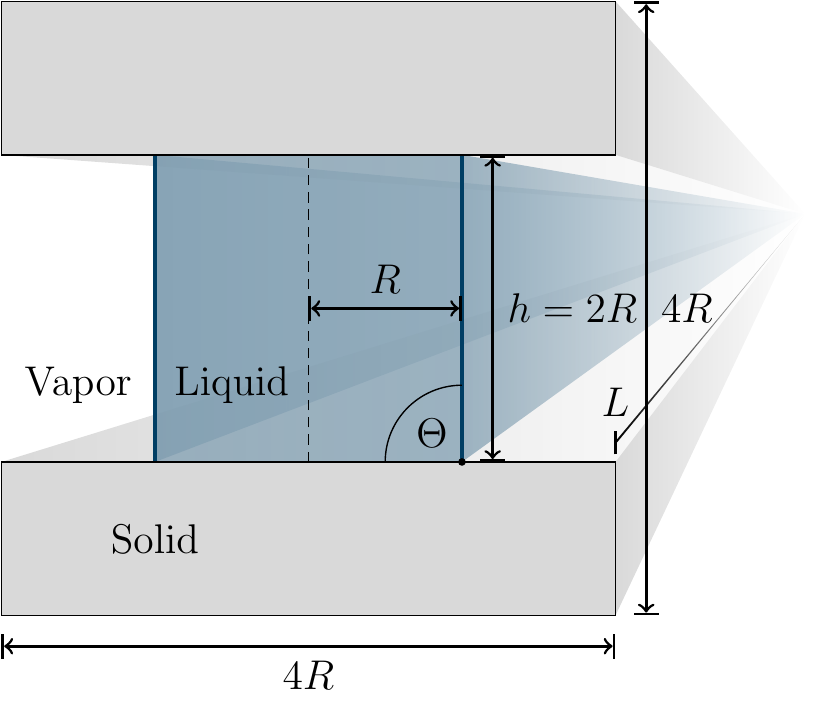}}
    \subfloat[\label{fig:plate_setup_cylinder}]{\includegraphics[width=0.5\linewidth]{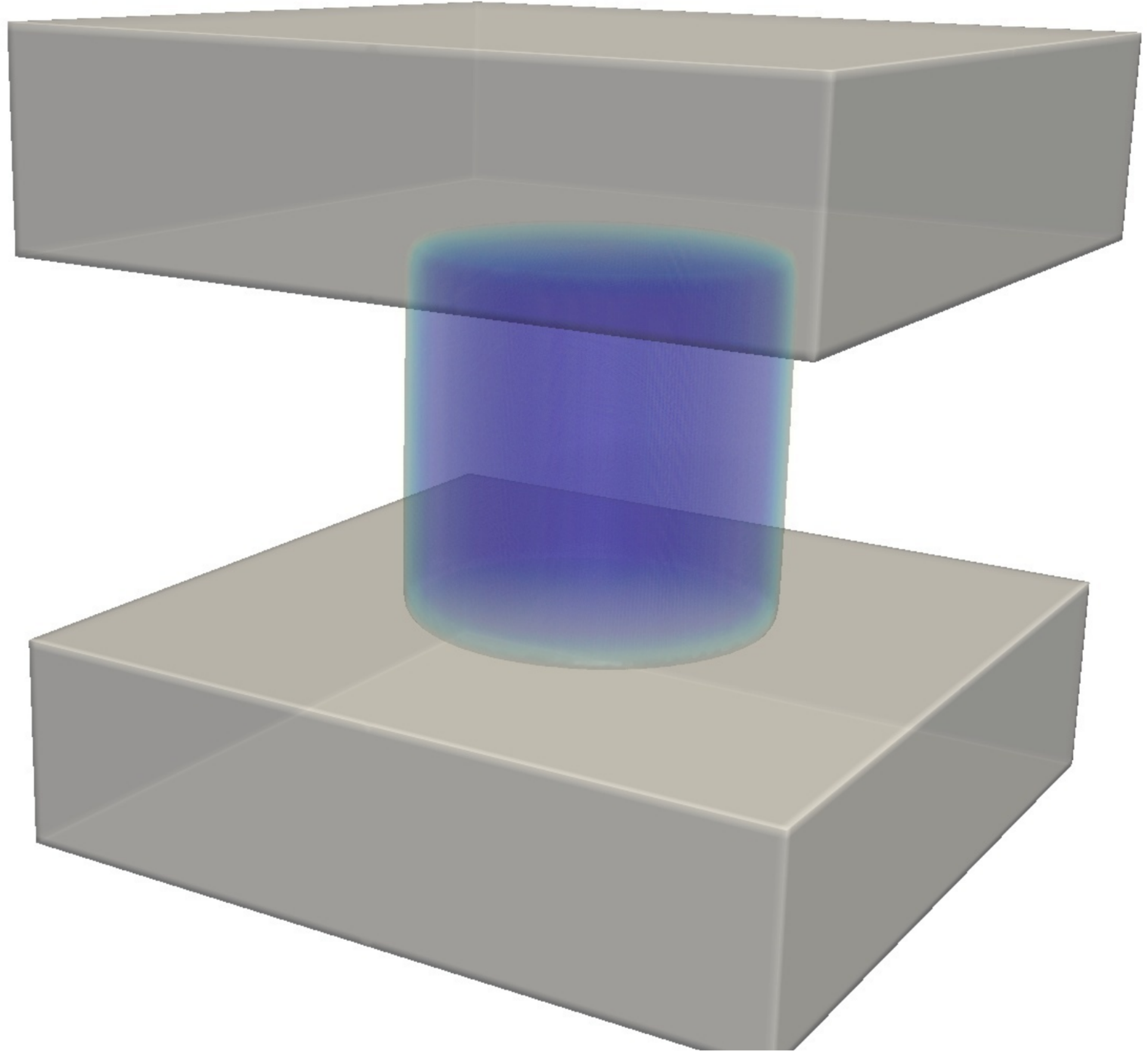}}
    \caption{(a) A schematic representation of a planar capillary bridge between two plates with the thickness \(R\), initial radius of the liquid bridge \(R\), and wetting angle \(\Theta = \pi/2\).
        (b) The initial simulation setup of a cylindrical capillary bridge.
        Depicted are the surfaces of the solid plates in gray and the liquid surface in blue.}
        \label{fig:plate_setup}
\end{figure}

We consider two scenarios as illustrated in Fig.~\ref{fig:plate_setup}.
The depicted capillary bridges are determined by their radius \(R\), the distance between the plates \(h\) and the wetting angle \(\Theta\).
To calculate the resulting force on the plates, we use the total differential of the internal energy \(U\) of such a capillary bridge which is given by

\begin{dmath} \label{eq:internal_energy_differential}
    dU =
    \sigma_{{SL}} \, dA_{{SL}}+
    \sigma_{{SV}} \, dA_{{SV}}+
    \sigma_{{LV}} \, dA_{{LV}}-
    p_{{V}}       \, dV_{{V}}-
    p_{{L}}       \, dV_{{L}}
\end{dmath}.
\(\sigma_{{SL}}\) denotes the surface tension of the solid-liquid interface, \(\sigma_{{SV}}\) of the solid-vapor interface, and \(\sigma_{LV}\) of the liquid-vapor interface with the associated areas \(A_{{SL}}\), \(A_{{SV}}\), and \(A_{{LV}}\).
\(p_{{V}}\) and \(p_{{L}}\) are pressures of the vapor and liquid phases with the respective volumes \(V_{{V}}\) and \(V_{{L}}\).
Because of the considered geometry, one can see that \(dA_{{SV}}=-dA_{{SL}}\) and \(dV_{{V}}=-dV_{{L}}\) so that Eq.~(\ref{eq:internal_energy_differential}) can be simplified

\begin{dmath} \label{eq:internal_energy_differential_simple}
    dU
    =
    - \left(\sigma_{{SV}} - \sigma_{{SL}}\right) \, dA_{{SL}}+
    \sigma_{{LV}} \, dA_{{LV}} -
    \Delta p \, dV_{{L}}
\end{dmath},
where \(\Delta p = p_{{L}} - p_{{V}}\) is the pressure difference between vapor and liquid.

The equilibrium shape of the liquid minimizes the internal energy \(U\).
In the absence of solids, in the case of a spherical liquid droplet (\(dA_{{SL}}=0\)), it can be shown by minimization of the total internal energy, i.e., by setting \(d U = 0\), that the pressure difference between vapor and liquid is proportional to the curvature \(\kappa\) of the surface,

\begin{dmath}
    \Delta{p} = \sigma_{{LV}} \kappa
    \label{eq:young_laplace}
\end{dmath},
which is called the Young-Laplace equation.
We consider here the curvature as the sum of the principal curvatures \(\kappa = \sum^{D-1}_i \kappa_i\) where \(D\) denotes the dimension.
Because a pressure gradient would result in a motion of the fluid, Eq.~(\ref{eq:young_laplace}) states that the equilibrium surface of the liquid must be of constant curvature.
In the presence of solid-fluid interfaces, the energy minimization principle delivers also, the well-known Young's equation
\begin{math}
    \cos{\Theta}
    = \left(\sigma_{{SV}}-\sigma_{{SL}}\right)/{\sigma_{{LV}}}
\end{math},
where \(\Theta\) is the wetting angle.

We consider a wetting of \(90^\circ\) so that the first term in Eq.~(\ref{eq:internal_energy_differential_simple}) vanishes.
Further, we neglect the effect of evaporation due to a change of liquid-vapor interface curvature so that the liquid volume can be considered as constant and hence the last term in Eq.~(\ref{eq:internal_energy_differential_simple}) vanishes also.
The change of the internal thus determined by the change of liquid-vapor interface area

\begin{dmath}
    dU
    =
    \sigma_{{LV}} \, dA_{{LV}}
\end{dmath}.

In the planar case (Fig.~\ref{fig:plate_setup_planar}), the magnitude of the force is then given by

\begin{dmath}\label{eq:plate_force_planar}
    F_{{p}} = 2 \sigma_{{LV}} L
\end{dmath},
where \(L\) is the length of the planar capillary wall.
Later, in the simulation, we use periodic boundary conditions.
\(L\) would then be the length of the simulation domain perpendicular to the depicted planar capillary bridge in Fig.~\ref{fig:plate_setup_planar}.

For the cylindrical capillary bridge (see Fig.~\ref{fig:plate_setup_cylinder}), one obtains

\begin{dmath}\label{eq:plate_force}
    F_{{c}} = \pi \sigma_{{LV}} R
\end{dmath}.
In the following, we drop the index of \(\sigma_{{LV}}\) to simplify the notation.
A more detailed derivation can be found in~\ref{app:theory_plate_force}.

\subsubsection[Motion]{Motion of the Plates}
\label{sec:theory_plate_motion}
Based on the analytic descriptions of the forces Eqs.~(\ref{eq:plate_force}) and~(\ref{eq:plate_force_planar}), the distance \(h\) of the plates as a function of time can be calculated.
This dynamic solution provides a more sophisticated benchmark for the dynamics of the present method.
To reduce the complexity of the analysis, we assume that the motion of the plates is slow compared to the motion of the fluid.
In this case, the bridge can be assumed to remain in its cylindrical shape during the motion of the plates.
Also, the kinetic energy of the liquid can be neglected.
These assumptions are valid if the mass of the plates is much larger than the mass of the liquid.

In the planar case, the computation is simple because the integral of the constant force Eq.~(\ref{eq:plate_force_planar}) delivers

\begin{dgroup}
\begin{dmath}
    h_{{p}} = h_0 - \frac{1}{2}\frac{F_{{p}}}{m} t^2 - \frac{1}{2}\frac{F_{{p}}}{m} t^2
\end{dmath}
\begin{dmath}
    h_{{p}}= h_0 - \frac{2\sigma L}{m} t^2
    \label{eq:theory_plates_motion_2d}
\end{dmath},
\end{dgroup}
where \(m\) is the mass of a single plate, \(h_0\) the initial distance, and \(t\) the time.
The factor of two accounts for the fact that both plates are mobile and subject to the same magnitude of the force.

In the cylindrical case, we consider  the inverse function \(h_{{c}} \left(t\right)\) can be approximated with

\begin{dmath}
    h_{{c}} \left(t\right) \approx h_0 - \frac{\pi \sigma R_0}{m} t^2
    \label{eq:theory_plate_motion_approx}
\end{dmath}
where the change of the capillary radius is neglected.

\subsection{Liquid Bridge Between two Spherical Solids}
\label{sec:theory_2_spheres}

As an introductive scenario, we consider a liquid bridge between two spherical solids of the same size in contact with each other.
Even in this simplified case, the analytic description of the bridge can be complicated.
The possible equilibrium configurations of such a system are determined by the wetting angle and the liquid volume fraction
\begin{math}
    c = V_{{L}}/\left(V_{{L}} + V_{{S}}\right)
\end{math},
where \(V_{{S}}\) is the solid volume.
The choice of these two parameters offers the possibility to discuss the topology of this system independent of its scale.
For two spherical solids in contact, two ideal solutions for the topology of the liquid bridge exist, a spherical bridge in the limit of a large liquid volume fraction \((c = 0.75,\Theta = 0)\) and a cylindrical bridge for the other extreme case of low liquid content \((c = 0.2,\Theta = 0)\).
Both solutions exist because they form a surface of constant curvature.
By calculating the wetting angle between a spherical liquid bridge as a function of the liquid volume fraction, one obtains the relation

\begin{dgroup}
    \begin{dmath}
        \Theta_{{S}}
        = \arccos\left(\frac{\chi_s^2-2}{2}\right)
        - \arccos\left(\frac{\chi_s}{2}\right)
    \end{dmath}
    \begin{dmath}
        \chi_s = \frac{2}{\sqrt[4]{3}} \frac{\sqrt{\left(1-c\right)\sqrt{c\left(1-c\right)}}}{1-c}
    \end{dmath}.
    \label{eq:theory_2_spheres_spherical}
\end{dgroup}

\begin{figure}
    \includegraphics[width = 0.618\linewidth]{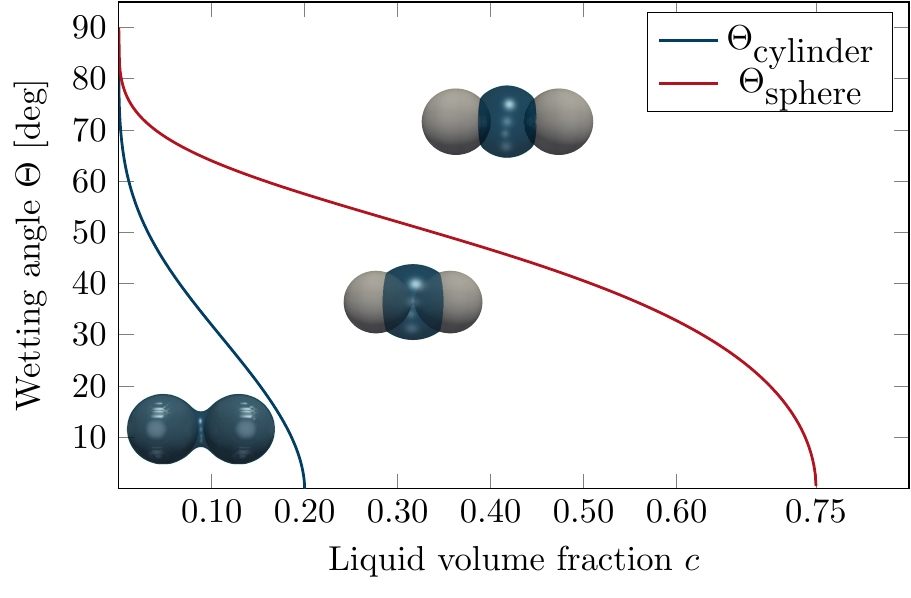}
    \caption{The wetting angle for the limiting cases of spherical and cylindrical liquid bridges as a function of the liquid volume fraction (see Eq.~(\ref{eq:theory_2_spheres_spherical}) and Eq.~(\ref{eq:theory_2_spheres_cylindrical})).
    The area between these two solutions corresponds to more complex (mixed) bridge shapes.
    These two functions thereby classify the topology of the bridges into three types as illustrated by pictures obtained from simulation.
    }
    \label{fig:theory_2spheres_phases}
\end{figure}

The same way, a relationship between the wetting angle and the liquid volume fraction can be found in the case of a cylindrical bridge,

\begin{dgroup}
    \begin{dmath}
        \Theta_{{c}}
        =
        \frac{\pi}{2} - \arccos \left(1 - \chi_c\right)
    \end{dmath}
    \begin{dmath}
        \chi_{{c}}
        =
        - \cos
        \left(
            \frac{1}{3} \,\arccos \left(\frac {1-9\,c}{1-c}\right)
            +\frac{\pi}{3}
        \right)
        + \frac{1}{2}
    \end{dmath}.
    \label{eq:theory_2_spheres_cylindrical}
\end{dgroup}
A derivation of Eq.~(\ref{eq:theory_2_spheres_spherical}) and Eq.~(\ref{eq:theory_2_spheres_cylindrical}) can be found in~\ref{app:theory_2_spheres}.
Equation.~(\ref{eq:theory_2_spheres_spherical}) and Eq.~(\ref{eq:theory_2_spheres_cylindrical}) are depicted in Fig.~\ref{fig:theory_2spheres_phases}.

However, more interesting are the parameter regions delimited by the spherical and cylindrical solutions.
For all pairs of wetting angles and liquid volume fraction below the cylindrical solution in Fig.~\ref{fig:theory_2spheres_phases}, one of the radii of curvature of the bridge becomes negative.
Between the cylindrical and the spherical solution lies the region of parameters where the bridges have a convex shape but not yet a spherical one.
In the region above the spherical solution, the spherical solids are not brought into contact by the liquid capillary bridge.
Instead, the solids are separated by the bridge.
As long as the solid surface is not entirely hydrophobic, the spherical solids will not separate from the liquid bridge.
Furthermore, the liquid bridge has a spherical shape for liquid volume fractions and wetting-angles above the plotted spherical solution.

Instead of determining the shape of the liquid bridge, the liquid fraction and the wetting angle determine the distance between the spherical solids.
In the following, we focus on the investigation of the parameter range below the spherical solution.
To illustrate these possible shapes, pictures obtained from simulation are added to Fig.~\ref{fig:theory_2spheres_phases} according to their contact angle and liquid volume fraction.
The pictures are obtained from simulations with the method described in Sec.~\ref{sec:model}.

\section{Model}
\label{sec:model}
In the following section, we give a brief compendium of the implemented model which consists basically of a combined approach of the multi-phase-field method for the modeling of solid phases, including solid-solid as well as solid-liquid phase transformation, and the Lattice Boltzmann method for modeling the fluid flow, liquid-vapor phase separation, and solid-fluid interaction.
An overview of the modeling of solids and their dynamics is shown in Sec.~\ref{sec:model_phase_field}.
The Lattice Boltzmann method is discussed in Sec.~\ref{sec:fluids} with its coupling to the dynamics of solid bodies.
In Sec.~\ref{sec:model_simulation_procedure}, we explore some numerical aspects and parameters.

\subsection{Modeling of Solids}
\label{sec:model_phase_field}
The phase-field method is a method for solving interfacial problems, and it has been applied to various kinds of problems such as solidification~\cite{steinbach_effect_2008}, grain-growth~\cite{darvishi_kamachali_3-d_2012}, surface or phase-boundary diffusion~\cite{schiedung_multi-phase-field_2017}, and elastic deformation of solid bodies due to surface tension~\cite{schiedung_multi-phase-field_2018}.
Here, we use the well-established multi-phase-field method which is summed up in two review articles~\cite{steinbach_phase-field_2009, steinbach_phase-field_2013}.

The model allows us to distinguish and track each solid particle while for example being deformed or being in the process of a phase transition.
This tracking of a rigid body and its topology is achieved by a continuous indicator function called the phase-field \(\phi_{\alpha}\).
Given a point \(\vec{x}\) in space,
\begin{math}
    \phi_{\alpha} \left(\vec{x}\right) = 1
\end{math}
means that the phase \(\alpha\) is present at that point and
\begin{math}
    \phi_{\alpha} \left(\vec{x}\right) = 0
\end{math}
means that it is not.
All values of \(\phi_{\alpha}\) between \(0\) and \(1\) are considered as surface or interface of the particle or phase \(\alpha\) with its surroundings.
The considered system may consist of \(N\) phases so that all phase-fields must satisfy the sum constraint
\begin{math}
    \sum_{\alpha}^{N} \phi_{\alpha} = 1
\end{math}.
A rigid body \(B\) must consist of at least one phase-field, but it may even consist of multiple phase-fields.
We consider the dynamics of the solid particles as modeled by the advection of the phase-fields according to the velocity of the solid bodies.
The velocity is calculated according to forces and torques acting on the bodies (see Sec.~\ref{sec:model_solid_dynamics}).
Between two colliding bodies, a repulsive force is used (see Sec.~\ref{sec:solid_solid_interaction}).

As mentioned above, the phase-field method is commonly used to describe phenomena such as phase transformations or even surface/phase-boundary diffusion which may lead to growth, shrinkage or deformation of solid particles.
These processes can be considered with the present model, but they are not considered in this work, and the reader is referred to~\cite{steinbach_phase-field_2009, steinbach_phase-field_2013, schiedung_multi-phase-field_2017}.

\subsubsection{Solid Dynamics}
\label{sec:model_solid_dynamics}
Consider a rigid body \(B\) which is subject to the force \(\vec{F}_{B}\) and the torque \(\vec{T}_{B}\).
The center of mass \(\vec{X}_{B}\) of the solid body is accelerated with

\begin{dmath}
    \ddot{\vec{X}}_{B} = M^{-1}_{B} \vec{F}_{B}
\end{dmath},
where \(M_B\) is the mass of the solid body.
The angular velocity \(\vec{\omega}_{B}\) changes with

\begin{dmath}
    \dot{\vec{\omega}}_{B}
    =
    \mat{I}^{-1}_{B} \vec{T}_{B}
\end{dmath},
where \(\mat{I}_{B}\) is the tensor of inertia.
The resulting velocity \(\vec{u}_{B}\) of \(B\) at the point \(\vec{x}\) can be calculated with

\begin{dmath}
    \vec{u}_{B} \left(\vec{x}\right)
    =
    \dot{\vec{X}}_{B} + \vec{\omega}_{B}
    \times\left(\vec{x} - \vec{X}_{B}\right)
    \label{eq:solid_velocity}
\end{dmath}.
Each of the phase-fields \(\phi_{\alpha}\) associated with the rigid body \(B\) is advected with

\begin{dmath}
    \dot{\phi}_{\alpha}
    =
    \vec{u}_{B}
    \cdot
    \nabla \phi_{\alpha}
    \label{eq:phase_advection}
\end{dmath}.
For the advection Eq.~(\ref{eq:phase_advection}) of the phase-fields, we use a high-order scheme with directional splitting and a monotonized central flux limiter~\cite{van_leer_towards_1977}.

\subsubsection{Solid-Solid Interactions}
\label{sec:solid_solid_interaction}
Collisions are nearly inevitable when simulating rigid bodies inside a dynamic fluid flow.
Thus, we utilize a repulsive solid-solid interaction to account for collisions.
This repulsive force between two rigid bodies is derived from a potential energy density \(e\left(\vec{x}_i,\vec{x}_j\right)\) between a point \(\vec{x}_i\) of the rigid body \(B_1\) and a point \(\vec{x}_i\) of the rigid body \(B_2\)

\begin{dmath}
    e\left(\vec{x}_i,\vec{x}_j\right)
    =
    e_0
    \left\{
    \begin{array}{ll}
        \left|\frac{\left\Vert \vec{x}_i-\vec{x}_j\right\Vert-r_c}{r_c}\right|^n & \textrm{if } \left\Vert \vec{x}_i - \vec{x}_j \right\Vert < r_c \\
        0 & \textrm{else}
    \end{array}
    \right.
    \label{eq:soild_solid_potential}
\end{dmath}
where \(r_c\) is a characteristic interaction distance and the energy density \(e_0\) determines the interactions strength.
Furthermore, a point \(\vec{x}\) is considered to be a part of the solid body \(B\) if one of its associated phase-fields is non zero.
The force density \(\vec{\mathsf f}_{ij}\) of a rigid body \(B_1\) at \(\vec{x}_j\) acting on a rigid body \(B_2\) at \(\vec{x}_i\) can obtain as

\begin{dgroup}
\begin{dmath}
    \vec{\mathsf f} \left(\vec{x}_i,\vec{x}_j \right)
    = \frac{\partial e\left(\vec{x}_i,\vec{x}_j \right)}{\partial \left(\vec{x}_i-\vec{x}_j\right)}
\end{dmath}
\begin{dmath}
    =
    e_0
    \frac{n}{r_c}
    \frac{\vec{x}_i-\vec{x}_j}{\left\Vert\vec{x}_i-\vec{x}_j\right\Vert}
    \left|\frac{\left\Vert \vec{x}_i-\vec{x}_j\right\Vert-r_c}{r_c}\right|^{n-1}
    \label{eq:solid_solid_force}
\end{dmath}.
\end{dgroup}
By the summation over all pairs of \(\vec{x}_i\) and \(\vec{x}_j\) within the characteristic interaction length, the resulting total forces on the rigid bodies can be calculated.
To keep the solid-solid interaction as short ranged as possible, we have chosen \(r_c = 3\,\Delta x\) and \(n=4\) in our simulations where \(\Delta x\) is the lattice spacing.

\subsection{Modeling of Fluids}
\label{sec:fluids}
We use for the simulation of fluids the lattice Boltzmann method which is summed up in Sec.~\ref{sec:lbm}.
The interaction of the fluid with the solid bodies is shown in Sec.~\ref{sec:solid_fluid_interaction}.
In Sec.~\ref{sec:force_benzi}, we introduce the implemented method for the liquid-vapor phase separation.

\subsubsection{The Lattice Boltzmann Method}
\label{sec:lbm}
Consider the distribution function \(f \left( \vec{x}, \vec{v}, t \right)\) which is the probability to find a pseudo particle at the position \(\vec{x}\) with the physical velocity \(\vec v\) at the time \(t\)~\cite{montessori_three-dimensional_2015}.
The continuous velocity space \(\vec{v}\) is discretized with a finite number of physical velocities \(\vec{c_i}\) of the pseudo particles.
Although many schemes have been proposed for the discretization of the velocity space, we only consider here the so-called three-dimensional twenty-seven velocity model (D3Q27) with

\begin{dmath}
    \vec{c}_i
    =
    \left\lbrace
    \begin{array}{ll}
        (0,0,0) & i = 0 \\
        (\pm c,0,0), (0,\pm c,0), (0,0,\pm c) & i = 1,\ldots,6 \\
        (\pm c,\pm c,0), (0,\pm c,\pm c), (\pm c,0,\pm c)  & i = 7,\ldots,18 \\
        (\pm c,\pm c,\pm c)& i = 19,\ldots26
    \end{array}
    \right.
\end{dmath}
where \(c=\Delta x/\Delta t\) is a shorthand and \(\Delta t\) is the time step.
The fluid density \(\rho\) is calculated as the sum of the distribution function overall velocities
\begin{math}
    \rho = \sum_i f_i
\end{math} where the shorthand
\begin{math}
    f_i \left( \vec{x}, t \right)
    \equiv f \left( \vec{x}, \vec{c}_i, t \right)
\end{math}
has been used.
Likewise, the fluid velocity \(\vec u\) is calculated as average overall physical velocities \(\vec{c_i}\) weighted by the amount of pseudo particles \(f_i\) with
\begin{math}
    \vec u = 1/{\rho} \sum_i \vec{c_i} f_i
\end{math}.
This way, the lattice Boltzmann equations can be written as

\begin{dgroup}
    \begin{dmath}
        f_{i}^* \left(\vec{x},t\right)
        =
        f_{i} \left(\vec{x},t\right)
        -\frac{1}{\tau} \left[ f_{i} \left(\vec{x},t\right) -f_{i}^{{eq}} \left(\rho\left(\vec{x},t\right), \vec{u}\left(\vec{x},t\right)\right)\right]
        +\Delta t F_i
        \label{eq:lb_collision}
    \end{dmath}
    \begin{dmath}
        f_{i} \left(\vec{x},t + \Delta t \right)
        =
        f_{i}^{*} \left(\vec{x}- \vec{c_i} \Delta{t},t\right)
        \label{eq:lb_streaming}
    \end{dmath}
    \label{eq:lattice_boltzmann}
\end{dgroup}
which describe the collision step Eq.~(\ref{eq:lb_collision}) and the streaming step Eq.~(\ref{eq:lb_streaming}) of the pseudoparticles, \(f_i\).
The right-hand side of Eq.~(\ref{eq:lb_collision}) consists of two parts, the relaxation of the distribution function towards the local equilibrium distribution \(f_{i}^{{eq}}\) and the forcing term \(F_i\).
The local equilibrium distribution function can be calculated with

\begin{dmath}\label{eq:distribution_eq}
    f_{i}^{{eq}} \left(\rho, \vec{u} \right)
    =
    \rho{}
    \sum_i w_i
    \left[1-
        \frac{\vec{u} \cdot \vec{u}}{2 c_s^2} +
        \frac{\vec{c_i} \cdot \vec{u}}{c_s^2} +
        \frac{\left(\vec{c_i} \cdot \vec{u}\right)^2}{2 c_s^4}
    \right]
\end{dmath},
which is a second-order expansion of the local Maxwell-Boltzmann distribution with the weights

\begin{dmath}
    w_i
    =
    \left\lbrace
    \begin{array}{ll}
        8/27  & i = 0 \\
        2/27  & i = 1,\ldots,6 \\
        1/54  & i = 7,\ldots,18 \\
        1/216 & i = 19,\ldots26
    \end{array}
    \right.
\end{dmath}.
\(\tau\) is the relaxation time and \(c_s\) the speed of sound in the lattice Boltzmann fluid with
\begin{math}
    c_s =c/\sqrt{3}
\end{math}.

We use the forcing term as published in~\cite{guo_discrete_2002} with

\begin{dmath}
    F_i
    =
    \left(1-\frac{1}{2\tau}\right) w_i \left[\frac{\vec{c}_i-\vec{v}}{c_s^2} + \frac{\vec{c}_i \cdot \vec{v}}{c_s^4} \vec{c}_i\right]
    \cdot
    \vec{\mathsf f}
    \label{eq:guo_forcing}
\end{dmath},
where \(\vec{\mathsf f}\) is the local body force density acting on the fluid and \(\vec{v}\) is the fluid velocity which can be calculated with

\begin{dmath}
    \rho\vec{v}
    =
    \sum_i \vec{c}_i f_i + \frac{\Delta t}{2} \vec{F}
    \label{eq:fluid_velocity}
\end{dmath}.
We apply body force densities originating from the solid-fluid interaction (Sec.~\ref{sec:solid_fluid_interaction}) and from non-ideal fluid contributions (Sec.~\ref{sec:force_benzi}) which cause a phase separation between the liquid and vapor phase.

\subsubsection{Solid-Fluid Interaction}
\label{sec:solid_fluid_interaction}
A rigid body and a fluid exert force on each other due to their motion and inertia.
In the following, we introduce so-called ``bounce-back'' method~\cite{aidun_lattice_1995} which has been implemented.

To use this solid-fluid interaction method a point in the simulation domain has to be either solid or liquid.
Because the phase-field method uses a continuous transition between two phases, a point is considered to be solid if the phase-field representing the fluid phase content is less then \(0.05\).
If a point and its adjacent points are considered to be fluid, the lattice method is applied (see Sec.\ref{sec:lbm}).
Near a solid surface, the bounce-back adds to the right-hand side of Eq.~(\ref{eq:lb_streaming}) and accounts for fluid which ``bounces back'' form the solid surface

\begin{dmath}
    \label{eq:lb_bounce_back}
    f_i \left(\vec{x}, t + \Delta t \right)
    =
    \left\{
    \begin{array}{ll}
        f^{*}_{-i} \left(\vec{x}, t\right) +
           g\left(\vec{x}, \vec{c_i}, \vec{u}_{B},t \right)
           & \textrm{if } \vec{x}-\vec{c_i}\Delta{t} \textrm{ is solid}
           \\
        f^{*}_i \left(\vec{x}-\vec{c_i}\Delta{t}, t\right)
           & \textrm{else}
    \end{array}
    \right..
\end{dmath}
If no point \(\vec{x}-\vec{c_i}\Delta{t}\) in the vicinity of \(\vec x\) is solid, the bounce-back is not calculated so that Eq.~(\ref{eq:lb_streaming}) and Eq.~(\ref{eq:lb_bounce_back}) deliver the same result.
However, if the point \(\vec{x}-\vec{c_i}\Delta t\) lies within a solid body, there is no fluid population that can be streamed to point \(\vec{x}\).
Hence, the pseudo particles coming from the point \(f_i\left(\vec{x}\right)\) with the velocity \(\vec{c_i}\) are assumed to ``bounce-back'' from the solid surface in the middle between the two points.
Consequently, \(f^{*}_{-i} \left(\vec{x}, t\right)\) denotes the population of the pseudo particles with the velocity \(-\vec{c}_i\).
If a solid surface is present in the vicinity, the corresponding part of the distribution function is reflected.
The solid surface is here assumed to be in the middle between the fluid and the rigid body node at \(\vec{x}-\vec{c_i}\Delta t /2\).
This adds to the consequence the remaining fluid has to be distributed to the neighboring points if a certain point changes from liquid to solid.

Further, we consider the movement of the solid body with the velocity \(\vec{u}_B\).
The movement of the solid body leads to a change of momentum of the adjacent fluid which is accounted for by

\begin{dmath}
    g \left(\vec{x}, \vec{c_i}, \vec{u}_{B} \left(\vec{x}\right) \right)
    =
    \frac{2\rho \left(\vec{x}\right)}{c_s^2}
    w_{i} \left.\vec{c_i} \cdot \vec{u}_{B}\left(\vec{x}-\vec{c}_i\Delta t /2\right)\right.
    + \delta \rho \left(\vec{x}\right) \delta_{i0}
    \label{eq:lb_bounce_back_us}
\end{dmath}
in Eq.~(\ref{eq:lb_bounce_back}).
Whereas \(\delta \rho \left(\vec{x}\right)\) is an auxiliary term that ensures the local mass conservation with

\begin{dmath}
    \label{eq:lb_bounce_back_mass}
    \delta \rho \left(\vec{x}\right) =
    \frac{2\rho \left(\vec{x}\right)}{c_s^2}
    w_{i}
    \vec{c_i} \cdot \vec{u}_{B}\left(\vec{x}-\vec{c}_i\Delta t /2\right)
\end{dmath},
where the sum over \(i\) is executed for each point \(\vec{x}-\vec{c}_i\Delta t\) that is solid so that the total change of mass caused by the first term in Eq.~(\ref{eq:lb_bounce_back_us}) is compensated.
This mass correction \(\delta \rho \left(\vec{x}\right)\) is then added to the resting fluid population with the term \(\delta_{i0}\) which is only non-zero for \(i=0\) so that no slip of the fluid along the solid surface is introduced.

The collision and bounce back of the fluid at the solid surface results in a momentum exchange between solid and fluid.
The resulting local change of the momentum density \(\Delta\vec{m}_B\) of the rigid body at a solid point \(\vec{x}\) is given by

\begin{dmath}
    \Delta\vec{m}_B \left(\vec{x}\right)
    =
    -
    \vec{c_i}
    \left[
        2 f^{*}_{i} \left(\vec{x}+\vec{c_i}\Delta{t}\right) +
        g\left(\vec{x}+\vec{c_i}\Delta{t}, \vec{c_i}, \vec{u}_{B} \right)
    \right]
\end{dmath}
where the sum is executed for each point \(\vec{x}+\vec{c}_i\Delta t\) that is fluid.
By integration over the entire solid body \(B\), corresponding total force \(\vec{F}_B\) and torque \(\vec{T}_B\) can then be obtained with

\begin{dgroup}
\begin{dmath}
    \vec{F}_B = \frac{1}{\Delta{t}}\int_{B}
    \Delta\vec{m}_B \left(\vec{x}\right){d} V
\end{dmath}
\begin{dmath}
    \vec{T}_B
    =
    \frac{1}{\Delta{t}}
    \int_{B}
    \left(\vec{x}-\vec{c_i}\frac{\Delta{t}}{2}-\vec{X}_B\right) \times
    \Delta\vec{m}_B \left(\vec{x}\right) {d} V
\end{dmath},
    \label{eq:solid_liquid_force}
\end{dgroup}
where \(\vec{X}_B\) denotes the center of mass of the solid body (see Sec.~\ref{sec:model_solid_dynamics}).

In addition to the introduced bounce-back, our simulations reveal that yet another solid-fluid interaction is necessary to conserve the fluid mass and momentum with a rigid body moving through it.
Consider a moving solid body in the discrete simulation space, a former fluid node at the point \(\vec{x}_n\) may become a solid node because a rigid body is passing by.
The remaining fluid distribution function \(f_i\left(\vec{x}_n\right)\) has to be distributed to the neighboring fluid points so that the fluid mass and momentum are conserved.
This conservation can be achieved with
\begin{dmath}
    f^{{new}}_i\left(\vec{x}\right) = f_i\left(\vec{x}\right) + \frac{f^{{eq}}_i\left( \rho\left(\vec{x_m}\right), \vec{u}\left(\vec{x_m}\right) \right)}{\left|N_F\left(\vec{x_m}\right)\right|}
\end{dmath},
where \(N_F\left(\vec{x}_n\right)\) is the set of fluid neighbor nodes in a sphere of the radius \(R_A\) around \(\vec{x}_n\) with
\begin{math}
    N_F\left(\vec{x_n}\right) = {\left.\left\{\vec{x}\,\right| \left\Vert \vec{x_n} - \vec{x} \right\Vert \leq R_A \textrm{~and }\vec{x} \neq \vec{x_n}\right\}}.
\end{math}
In our simulations, the choice of \(R_A = \Delta x\) has shown to be sufficient.

In the case that a former solid node becomes fluid, it is sufficient to set the fluid density to zero.

\subsubsection{Non-Ideal Fluids}
\label{sec:force_benzi}
In the following, we briefly present the implemented method to simulate liquid and vapor phase separation and their interaction with solid surfaces.
For more detailed information, the reader is referred to~\cite{shan_lattice_1993,shan_simulation_1994,montessori_three-dimensional_2015,benzi_mesoscopic_2006}.
The liquid and vapor phase separation is introduced by a force density \(\vec{\mathsf f}\) acting on the fluid (see Eq.~(\ref{eq:lattice_boltzmann}) and~(\ref{eq:guo_forcing})) with

\begin{dmath}\label{eq:benzi_force}
    \vec{\mathsf f} \left(\vec{x}\right) = -
    \mathcal{G} \psi{\left(\vec{x}\right)}
    \sum_{i} w_i \psi{\left(\vec{x} + \vec{c_{i}} \Delta{}t \right)} \vec{c_{i}}
\end{dmath},
where \(\mathcal G\) is a coupling constant and \(\psi\) a so-called lattice version of the mean-field potential,

\begin{dmath}
    \psi\left(\vec{x}\right) = 1 - e^{-\frac{\rho\left(\vec{x}\right)}{\rho_0}}
    \label{eq:mean_field_potential}
\end{dmath},
where \(\rho_0\) is a reference density.
The force density Eq.~(\ref{eq:benzi_force}) can be interpreted as the sum of force densities which act between the pseudo particles at \(\vec{x}\) and the surrounding pseudo particles at \(\vec{x} + \vec{c_{i}} \Delta{}t\).
Within this model, a convenient way to account for wetting effects is to identify a solid node as an immobile ``fluid'' with a fictions density \(\varrho_B\).
By the integration over all force densities between a fluid point \(\vec{x}\) and an adjacent point \(\vec{x} + \vec{c_{i}}\Delta{}t\) of the rigid body \(B\), the resulting force \(\vec{F}_B\) and torque \(\vec{T}_B\) acting on it can then be obtained

\begin{dgroup}
\begin{dmath}
    \vec{F}_B
    = -
    \mathcal{G}
    \int_F d\vec{x} \,
    \psi{\left(\vec{x}\right)}
    \,\,
    w_i \psi{\left(\vec{x} + \vec{c_{i}} \Delta{}t \right)} \vec{c_{i}}
\end{dmath}
\begin{dmath}
    \vec{T}_B =
    -\mathcal{G}
    \int_F d\vec{x} \,
    \psi{\left(\vec{x}\right)}
    \,\,
    \left( \vec{x} + \vec{c_i} - \vec{X}_B \right) \times
    w_i \psi{\left(\varrho_B \right)} \vec{c_{i}}
\end{dmath}
\end{dgroup}
where again the sum over \(i\) is executed for each point \(\vec{x}+\vec{c}_i\Delta t\) that is solid.

With
\begin{math}
    \vec{\mathsf f} = - \nabla \cdot \mat p
\end{math}
it is possible to identify the corresponding pressure tensor

\begin{dmath}
    p_{ij} =
    \left.\left[
        c_s^2 \rho +
        \frac{1}{2} c_s^2 \mathcal{G} \psi^2 +
        \frac{1}{2} c_s^4 \mathcal{G} \psi\Delta\psi +
        \frac{\mathcal{G} c_s^4}{4} \left\vert\nabla\psi\right\vert^2
    \right] \delta_{ij} \right.
    - \frac{1}{2} c_s^4 \mathcal{G} \partial_i \psi\partial_j \psi
    \label{eq:pressure_tensor}
\end{dmath}.
For the bulk pressure or the equation of states, one thus has

\begin{dmath}
    p = c_s^2\rho + \frac{1}{2} c_s^2 \mathcal{G} \psi^2
    \label{eq:pressure}
\end{dmath}.
The pressure tensor can be used to calculate the surface tension.
According to~\cite{rowlinson_molecular_1982}, the interface tension \(\sigma\) is given by the difference of the normal component of the pressure tensor \(p_\bot\) and its tangential component \(p_\parallel\) across the interface with

\begin{dmath}\label{eq:def_surface_tension}
    \sigma= \int_0^\eta {d}n
    \left(p_{\bot}- p_{\parallel}\right)
\end{dmath},
where \(n\) is the direction normal to the interface.
If one considers only a planar interface, so that \(\psi\) changes only along the normal direction of the interface, the interface tension can be calculated with

\begin{dmath}
    \sigma=-
    \frac{1}{2} c_x^4 \mathcal{G} \int\left\vert\frac{\partial\psi}{\partial{n}}
    \right\vert^2 \, {d} n
    \label{eq:interface_tension}
\end{dmath},
which is obtained by inserting Eq.~(\ref{eq:pressure_tensor}) into Eq.~(\ref{eq:def_surface_tension}).

Considering Eq.~(\ref{eq:interface_tension}) and Eq.~(\ref{eq:mean_field_potential}), one can see that the interface tension depends on the respective density profile.
The density profile across the liquid-vapor interface is determined by the coupling constant \(\mathcal{G}\).
The density profile across the solid-vapor and the solid-liquid can be adjusted tuning the above introduced ``solid density'' \(\varrho_B\).
This way, the contact angle of the solid body can be adjusted, and since the term ``solid density'' is misleading in this context, we refer to \(\varrho_B\) as the wetting parameter.
For a more detailed explanation, we refer the reader to~\cite{benzi_mesoscopic_2006}.

\subsection{Simulation Procedure}
\label{sec:model_simulation_procedure}
As mention in Sec.~\ref{sec:model_solid_dynamics}, we use a high-resolution scheme for the advection, Eq.~(\ref{eq:phase_advection}), of the phase-fields with directional splitting and a monotonized central flux limiter~\cite{van_leer_towards_1977}.
This higher-order scheme serves to reduce the well-known numerical problem of diffuse interface spreading under advection.
An advantage of the multi-phase-field method proposed here is that accounting for phase-field kinetics stabilizes the interface profile further, thus improving numerical stability.
As known from the standard literature on multi-phase-field~\cite{steinbach_phase-field_2009,steinbach_phase-field_2013}, the kinetics of \(\phi\) read,

\begin{dgroup}
\begin{dmath}
    \dot{\phi_\alpha}
    =
    \sum_{\beta \neq \alpha}^N
    \frac{M_{\alpha\beta}}{N}
    \left(
    \left[
        \sigma_{\alpha\beta} \left( I_\alpha - I_\beta \right) +
        \sum_{\gamma \neq \alpha,\beta} \left(\sigma_{\beta\gamma} - \sigma_{\alpha\gamma}\right) I_{\gamma}
    \right] +
    \frac{\pi^2}{4\eta} \Delta g_{\alpha\beta}
    \right)
    \label{eq:sim_proc_phase_field_dynamics}
\end{dmath}
\begin{dmath}
    I_\alpha = \nabla^2\phi_\alpha + \frac{\pi^2}{\eta^2}\phi_\alpha,
\end{dmath}
\end{dgroup}
where \(M_{\alpha\beta}\) is the mobility of the interface between phases \(\alpha\) and \(\beta\), \(\sigma_{\alpha\beta}\) is the interface energy between the phase-fields and \(\eta\) is the interface width (see~\cite{steinbach_phase-field_2013} for more details).

The restoration of the phase-field profile comes with a trade-off.
Eq.~(\ref{eq:sim_proc_phase_field_dynamics}) restores the phase-field profile but also introduces a shrinkage of the phase-field

\begin{dmath}
    \Delta \phi_\alpha = \int \left[{\phi_\alpha\left(t_{{ref}}\right)} - {\phi_\alpha\left(t\right)}\right] \,dV
\end{dmath}
according to the local curvature.
Because of this shrinkage, the last term \(\Delta g_{\alpha\beta}\) is necessary which is dynamically determined to conserve the volume of a phase-field \(\phi_\alpha\) with

\begin{dmath}
    {\Delta g_{\alpha\beta} = - \Delta g_{\beta\alpha}} = \frac{4\eta}{\pi^2} \frac{\Delta \phi_\alpha}{\left|\delta \phi_\alpha \right|}
\end{dmath}.
Similar as in Sec.~\ref{sec:solid_solid_interaction}, \(\left|\delta \phi_\alpha \right|\) denotes the number of nodes in the interface region of \(\phi_\alpha\) with \(\delta \phi_\alpha = \left.\left\{ \vec{x} \right| 0.0 < {\phi\left(\vec{x}\right)} < 1.0\right\}\).
To minimize this solid-phase sintering, we chose the phase-field mobility as small as reasonable.

If not stated otherwise, the following values of the introduced parameters have been used for the present simulations:
\begin{table}[h]
    \begin{tabular}{l c r}
        \hline\hline
        %
        \(\rho_0\)           & reference density                    &  1.000 \\
        \(\rho_L\)           & liquid density                       &  1.888 \\
        \(\tau\)             & relaxation time                      &  1.100 \\
        \(\mathcal{G}\)      & coupling constant                    & -5.000 \\
        \(\rho_{\alpha}\)    & mass density of \(\phi_\alpha\)      & various\\
        \(\varrho_{\alpha}\) & wetting parameter of \(\phi_\alpha\) &  2.000 \\

        \(M_{\alpha\beta}\)  & phase-field mobility                 & \(10^{-5}\) \\
        \(\eta\)             & phase-field interface width          &  5.000 \\
        \hline\hline
    \end{tabular}
\end{table}

\section{Results}
\label{sec:results}
We first investigate the accuracy of the surface tension calculation in Sec.~\ref{sec:surface_tenison}.
This investigation is the basis for the testing of the present model.
In Sec.~\ref{sec:results_plate}, we test our model by the simulation of capillary bridges between two finite parallel plates.
The results obtained from these simulations are compared with analytic solutions given in section~\ref{sec:theory_plate_force}.
To demonstrate the capability to model the process of liquid-phase-sintering,
we also investigate a more complicated situation in Sec.~\ref{sec:results_dynamics}, the dynamics of multiple particles connected by a capillary bridge.
This investigation provides a first step towards the modeling of larger and more complex systems with more solid particles.

\subsection{Surface Tension}
\label{sec:surface_tenison}
\begin{figure}
    \includegraphics[width=0.618\linewidth]{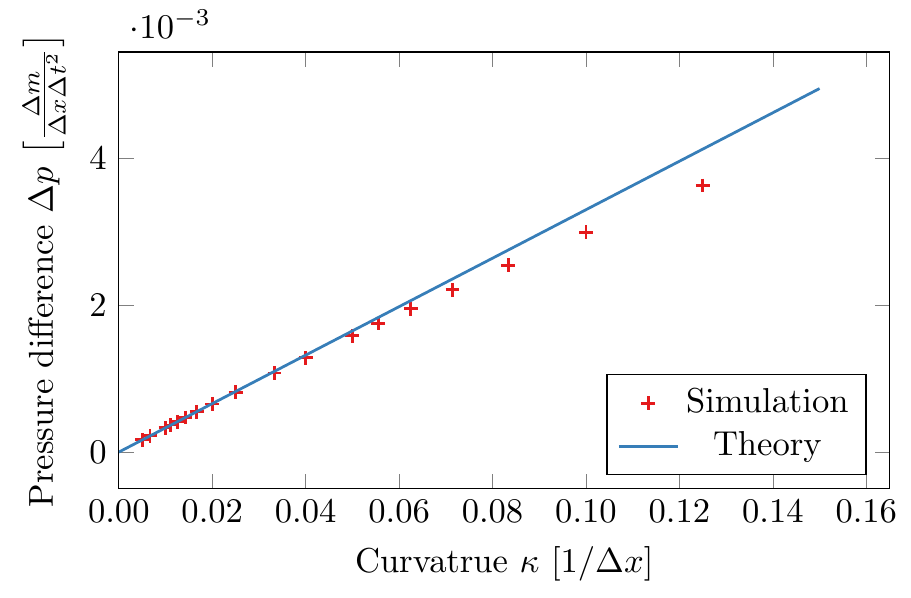}
    \caption{The figure shows the pressure difference between the vapor and the inside of a cylindrical liquid droplet for different radii which have been calculated with Eq.~(\ref{eq:pressure}).
    The solid line shows the expected Laplace pressure Eq.~(\ref{eq:young_laplace}) for a surface tension of \(3.297 \cdot 10^{-2}\).
    This value of the surface tension has been calculated with Eq.~(\ref{eq:interface_tension}).}
    \label{fig:lb2phase}
\end{figure}

In the following, we test the consistency of the surface tension calculations Eq.~(\ref{eq:interface_tension}) with the Laplace pressure Eq.~(\ref{eq:young_laplace}).
By evaluating Eq.~(\ref{eq:interface_tension}) for droplets of different radii, we obtained the value \(3.297 \cdot 10^{-2}\) in lattice units for the surface tension.
The value is used to predict the pressure gradient, by inserting it into Eq.~(\ref{eq:young_laplace}).
We show this prediction in Fig.~\ref{fig:lb2phase}.
Also, the differences for different radii, obtained with Eq.~(\ref{eq:pressure}), are depicted.
One can see that the obtained pressure gradients are in agreement with the ones predicted by the Laplace pressure, Eq.~(\ref{eq:young_laplace}).
A deviation from the predicted value can only be seen for large curvatures, which is expected because the spatial resolution decreases.

Another possibility to obtain the surface tension is to calculate the pressure gradient with Eq.~(\ref{eq:pressure}) and use it together with the radius as input for Eq.~(\ref{eq:young_laplace}) to calculate the resulting surface tension.
This way, we yield \(3.583 \cdot 10^{-2}\) for the value of surface tension which differs from the value obtained with Eq.~(\ref{eq:interface_tension}) by approximately \(8\%\).
In the following chapter, we show that this value for the surface tension provides a better estimate for the acting capillary force on a solid body.
Hence, \(\sigma = 3.583 \cdot 10^{-2}\) is used in the following.

\subsection{Capillary Bridge Between Two Plates}
\label{sec:results_plate}
We continue the testing of the present model by investigating a capillary bridge between two finite parallel plates as schematically depicted in Fig.~\ref{fig:plate_setup}.
The choice of finite plates ensures that the ambient pressure on both sides of the plate is the same.
We use the analytic solution to approximate the resulting motion of the plates and compare them with the simulation results.

\label{sec:results_plate_motion}

\begin{figure*}
    \subfloat[\label{fig:plate_motion_2d}]{\includegraphics[width=0.5\linewidth]{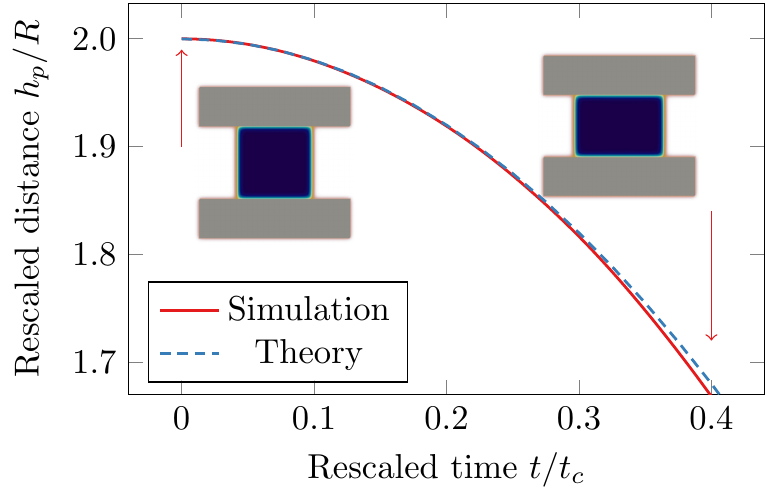}}
    \subfloat[\label{fig:plate_motion_3d}]{\includegraphics[width=0.5\linewidth]{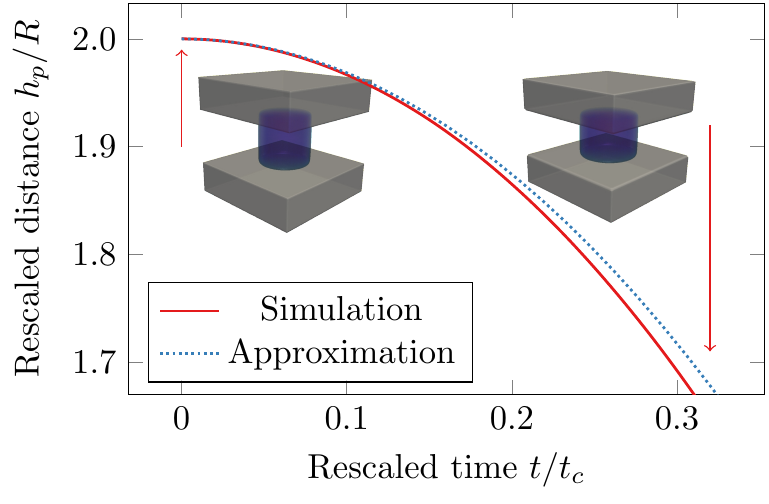}}
    \caption{The temporal evolution of the distance between two plates connected by (a) a planar capillary bridge and (b) a cylindrical capillary bridge with a \(90^\circ\) wetting angle are shown.
    The distance \(h\) of the plates has been rescaled by the length \(R\) which is the initial radius of the cylindrical capillary bridge.
    Furthermore, we have rescaled the time by a characteristic time \(t_c\) with \(t_c= \sqrt{m/\sigma}\) where \(m\) is the mass of the solid plates.
    For comparison, we show the theoretical expectations, Eq.~(\ref{eq:plate_motion_rescaled_planar}) in (a) and the approximation Eq.~(\ref{eq:plate_motion_rescaled_cylindrical_approx}).
    The simulation setups at \(t=0\) and the end of the displayed time interval are shown as insets.
    In both cases, the spatial discretization is the same with \(\Delta x/R= 0.05 \).
    }
    \label{fig:plate_motion}
\end{figure*}

In Sec.~\ref{sec:theory_plate_motion}, we introduced the approximative solution Eq.~(\ref{eq:theory_plate_motion_approx}) for the motion of the plates connected by a cylindrical capillary bridge and the analytic solution Eq.~(\ref{eq:theory_plates_motion_2d}) for the simple planar case.
These solutions are a good approximation if the solid motion is slow and the fluid is near its equilibrium state.
To realize this assumption in the simulation, we assign to the solid a high mass density \(\rho_{S}\) as compared to the liquid mass density \(\rho_L\) with \(\rho_{S}/\rho_{L} = 1.72 \cdot 10^4\).
The initial simulation setup is schematically depicted for the planar setup in Fig.~\ref{fig:plate_setup_planar} and for the cylindrical setup in Fig.~\ref{fig:plate_setup_cylinder}.
Figure~\ref{fig:plate_motion} shows the decreasing distance of the plates over time due to the action of the capillary force.
The distance of the plates has been rescaled by the length \(R\) which is the initial radius of the cylindrical capillary bridge.
Furthermore, we have rescaled the time by a characteristic time \(t_c\) with \(t_c= \sqrt{m/\sigma}\) where \(m\) is the mass of the solid plates.
By using \(R\) and \(t\), one can rewrite Eqs.~(\ref{eq:theory_plates_motion_2d}) and~(\ref{eq:theory_plate_motion_approx})

\begin{dgroup}
\begin{dmath}
    \frac{h_p}{R} = \frac{h_0}{R} - 2\left(\frac{t}{t_c}\right)^2
    \label{eq:plate_motion_rescaled_planar}
\end{dmath}
\begin{dmath}
    \frac{h_c}{R} \approx \frac{h_0}{R} - \pi \left(\frac{t}{t_c}\right)^2
    \label{eq:plate_motion_rescaled_cylindrical_approx}
\end{dmath}.
    \label{eq:plate_motion_rescaled}
\end{dgroup}
To obtain the rescaled equations of motion Eq.~(\ref{eq:plate_motion_rescaled}), we have assumed for the length of the planar capillary bridge that \(L=R\).
The simulation results are depicted in Fig.~\ref{fig:plate_motion} together with the rescaled equations of motion Eq.~(\ref{eq:plate_motion_rescaled}).
We also show as insets in Fig.~\ref{fig:plate_motion} the particular simulation setups at \(t=0\) and at the end of the displayed time interval.
A deviation of the simulations from the predictions is expected because the fluid motion has been neglected in the derivation of Eq.~(\ref{eq:plate_motion_rescaled}).
Consequently, we show only the very beginning of the motion in Fig.~\ref{fig:plate_motion} so that the theoretical prediction is accurate and can be used to test the simulations.
One can see that in both cases, the planar one in Fig.~\ref{fig:plate_motion_2d} and the cylindrical one Fig.~\ref{fig:plate_motion_3d}, the results agree with the predictions of Eqs.~(\ref{eq:plate_motion_rescaled_planar}) and~(\ref{eq:plate_motion_rescaled_cylindrical_approx}) in the beginning.
Nevertheless, for later times the simulation results deviate from the predictions.
This effect is more significant in the cylindrical case than in the planar one.

\subsection{Dynamics of Multiple Particles}
\label{sec:results_dynamics}

\subsubsection{Two Particles Connected by a Capillary Bridge}
\label{sec:results_dynamics_two}

\begin{figure*}
    \centering
    \subfloat[\label{fig:2spheres_time_velocity_begin}]{\includegraphics[width = 0.191\linewidth]{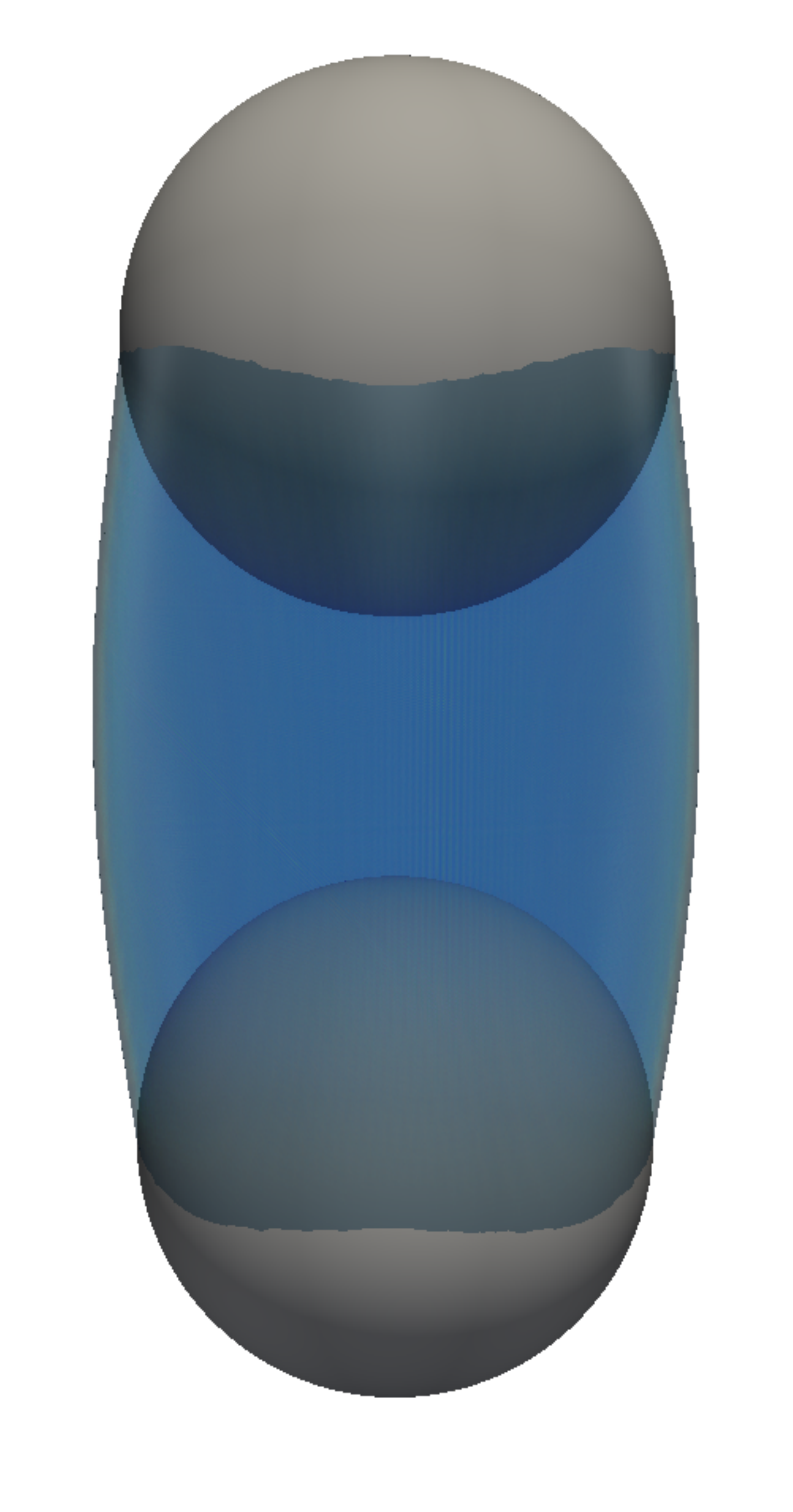}}
    \subfloat[                                        ]{\includegraphics[width = 0.618\linewidth]{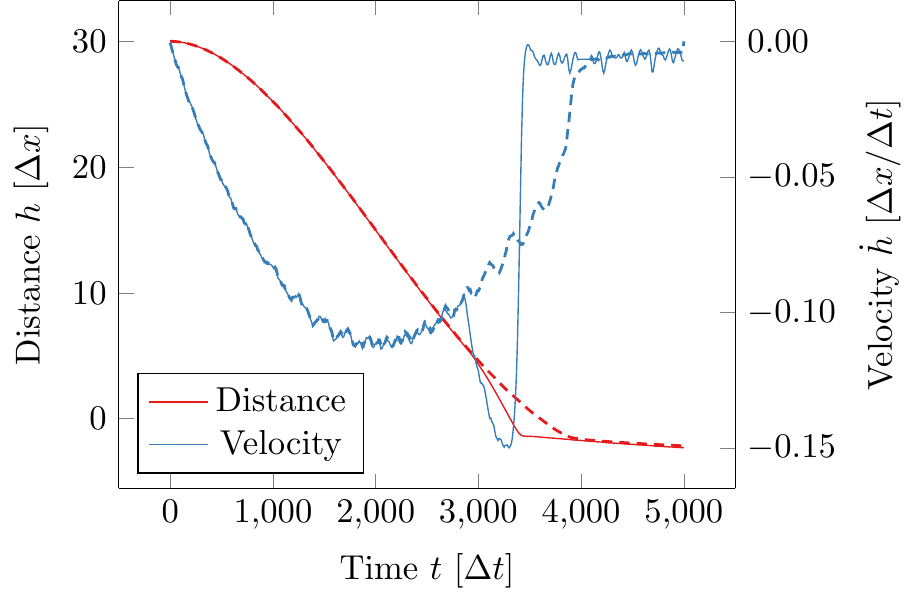}}
    \subfloat[                                        ]{\includegraphics[width = 0.191\linewidth]{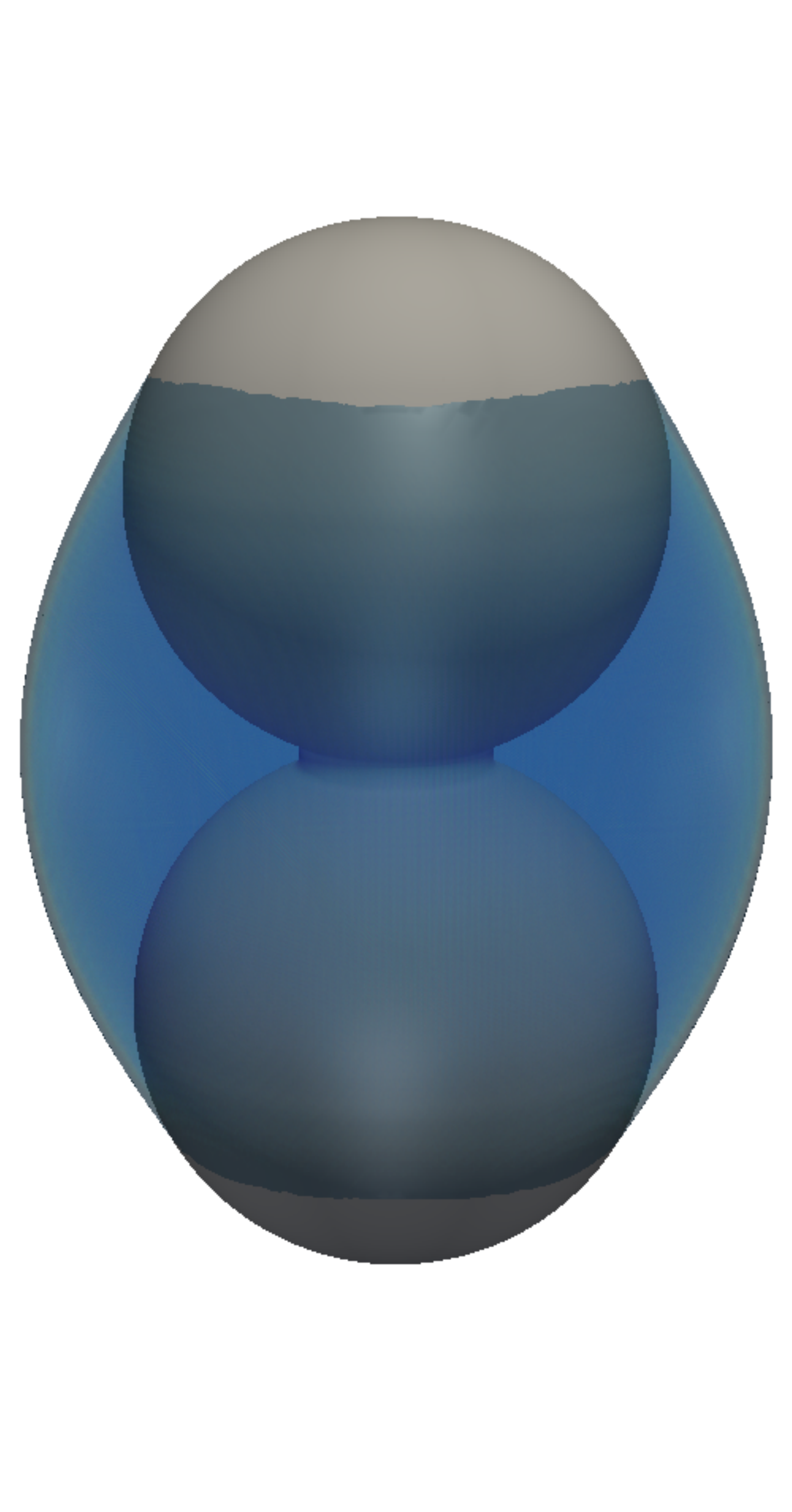}}
    \caption{
        (a) shows the simulation setup at \(t=0\) of the two solid particles with the radius \(R = 30\) and the liquid-vapor interface, represented by the blue transparent isosurface.
        The amount of liquid has been chosen in such a way that the liquid volume fraction is exact \(c=0.5\) at \(t=0\).
        (b) shows the closest distance \(h\) between the surfaces of the two spherical particles and the corresponding velocity \(\dot{h}\).
        Also, the same simulation is shown as a dashed line but without the phase-field dynamics so that the solids do not undergo sintering.
        The steep change in velocity (solid blue line) is thus the result of sintering processes, which sets on as soon as the surfaces of particles come into contact.
        (c) shows the simulation setup at \(t=5000\).
    }
    \label{fig:2spheres_time_velocity}
\end{figure*}

We continue our investigation with two spherical particles connected by a capillary bridge with a particle radius of \(R = 30\), mass density of \(\rho = 4\), and liquid fraction of \(c=0.5\) (see Fig.~\ref{fig:2spheres_time_velocity}).
The closest distance between the particles \(h\) and the velocity \(\dot{h}\) are shown in Fig.~\ref{fig:2spheres_time_velocity}.
Because we want to consider solids with a very high wettability, a wetting-parameter of \(\varrho_B = 2.0\) has been chosen.
Above in section~\ref{sec:model_simulation_procedure}, we have introduced the phase-field dynamics Eq.~(\ref{eq:sim_proc_phase_field_dynamics}) which models the physical process of sintering.
To show the influence of the phase-field dynamics, the simulation has been performed twice, once with the phase-field dynamics and once without it where the phase-fields are only advected.

In Fig.~\ref{fig:2spheres_time_velocity}, the closest distance between the surfaces of the particles \(h\) and the corresponding velocity \(\dot{h}\) are shown.
The dashed lines in Fig~\ref{fig:2spheres_time_velocity} indicate the simulations where the phase-fields are only advected.
Without the phase-field dynamics, it takes more time for the two particles to come into contact with each other.
The reason for this can be seen when looking at the velocity profile without the phase-field dynamics.
In the beginning, the two particles are strongly accelerated towards each other, until about two thousand time steps when the velocity reaches its peak value (velocity minimum of the dashed line in Fig.~\ref{fig:2spheres_time_velocity}).
After two-thousand time-steps, the two particles are continuously decelerated until the particles come into contact.
By looking at Fig.~\ref{fig:theory_2spheres_phases}, one can see that there exists an equilibrium solution for \(c=0.5\) and low wetting angles where the particles are in contact.
The existence of this equilibrium solution means that the capillary force must be attractive during the whole simulation time.
Furthermore, the phase-fields are purely advected, and the solid-solid interaction has only a range of 3.
Hence, the observed deceleration in Fig.~\ref{fig:2spheres_time_velocity} can only be caused by the dynamics of the fluid.
We do not consider an explicit drag force here.
Consequently, the bounce-back effect Eq.~(\ref{eq:lb_bounce_back}) can only cause this deceleration.

When the phase-fields are not only advected (solid lines in Fig~\ref{fig:2spheres_time_velocity}), one can see that the time, until the two particles come into contact, is reduced.
Even more visible is the influence of the phase-field dynamics in the velocity profile.
At about \(t=3000\), the magnitude of the velocity increases strongly, and a short time after that the velocity drops to nearly zero.
This strong acceleration occurs at about a distance of \(10\) and can be understood in the context of the phase-field method.
The value of the present phase-fields varies continuously from \(0\) to \(1\) within a distance \(\eta = 5\).
This defines the interface thickness.
Hence, the phase-fields begin to touch each other about a distance of \(2 \eta = 10\).
At this distance, the phase-fields start to merge and form a common interface between the two solid particles.

When \(\eta\) is comparable to the width of a grain-boundary, the length-scale can be seen as physically meaningful, and the described process of merging can be regarded as the first step in the solid-phase sintering process.
In contrast, if \(\eta\) is larger than a physical grain boundary width, the dynamics of the phase-field and the resulting variation of the distance below \(2\eta\) shall be viewed as a numerical feature to model solid-phase sintering of larger scales.

At a distance below \(2\eta\), the phase-fields representing the particles start to deform.
This occurs due to the solid-solid sintering process, where the particles reduce the surface area at the expense of a larger grain-boundary area.
(The grain-boundary energy is assumed to be smaller than the surface energy).
During this process, the centers of mass come closer together, as compared to hard and inert bodies at contact.
This feature shows itself as a small negative distance in Fig.~\ref{fig:2spheres_time_velocity}.

For more details about the modeling of surface diffusion with phase-field and the evolution of stress in two spherical particles undergoing the process of solid-phase sintering, the reader is referred to~\cite{schiedung_multi-phase-field_2017,kundin_phase-field_2018,schiedung_multi-phase-field_2018}.

\subsubsection{Effect of Liquid Fraction}
\label{sec:results_dynamics_c}

Above, we investigated the dynamics of a capillary bridge between two spherical particles.
A more intriguing investigation, however, is the influence of the liquid fraction on the dynamics of the system.
The influence of the liquid fraction on the topology of a liquid bridge between two spheres has been shown above in Sec.~\ref{sec:theory_2_spheres}.
\begin{figure*}
    \centering
    \includegraphics[width = \linewidth]{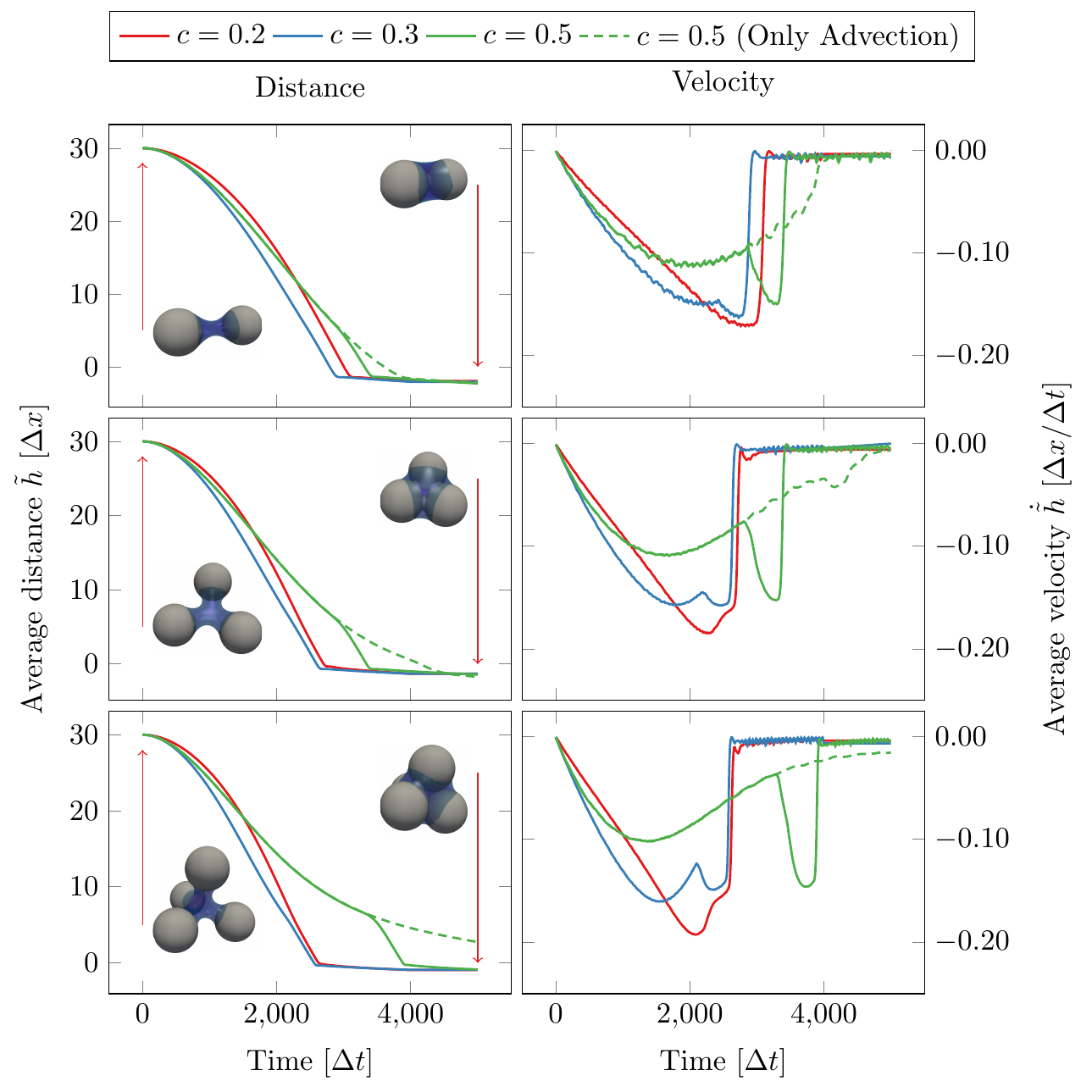}
    \caption{In the first row, the distance (left) and the velocity (right) of two spherical particles are shown which have the radius \(R = 30\) and are connected by a capillary bridge.
    The amounts of liquid are chosen in such a way that at \(t=0\) three different liquid to solid volume fractions are investigated, \(c=0.2\), \(c=0.3\), and \(c=0.5\).
    As inset picture, we show the configurations of the liquid and the solid particles at \(t=0\) and \(t=5000\) for a liquid fraction \(c=0.2\).
    Likewise, the second row shows the average distance and velocity of three particles forming an equilateral triangle.
    In the third raw the results of four-particles forming a tetrahedron are shown.
    }
    \label{fig:spheres}
\end{figure*}

We start with two spherical particles with a radius of \(R = 30\) and a mass density of \(\rho = 4\) which are initialized with different amounts of liquid between them.
We continue with the investigation of three and four particles, as a forecast of a system with many particles.
Similar to section~\ref{sec:results_dynamics_two}, before the particles are allowed to move, the simulations run for fifteen thousand simulation time steps so that the liquid reaches its equilibrium shape.
The time \(t=0\) refers to the point in time when the particles are allowed to move.
Furthermore, we investigate here three different liquid fraction \(c=0.2\), \(c=0.3\), and \(c=0.5\).
The simulation setup for \(c=0.2\) at \(t=0\) and \(t=5000\) are depicted as insets in the corresponding plots in Fig.~\ref{fig:spheres}.
For a liquid fraction of \(c=0.5\), we show again with a dashed line for a simulation where the phase-field dynamics, Eq.~(\ref{eq:sim_proc_phase_field_dynamics}), is switched off so that the phase-fields are purely advected.

We start the discussion with the concentration dependency of the distance and velocity profiles of two particles.
In the case of a small amount of liquid (\(c=0.2\)), the velocity of the particles increases linearly with time until they come into contact.
For a liquid fraction of \(c=0.3\), one can see an increase in the velocity of the particles at the beginning, compared to the case of \(c=0.2\).
By looking at the velocity profile of \(c=0.5\), the particles are more strongly accelerated at the beginning than for \(c=0.2\), but for later times one can see a clear deceleration.
To show this more clearly, we also plot the velocity profile corresponding to the case where the phase-fields are purely advected (dashed-line).
By comparing the dashed and the solid lines of the velocity profile, one can see that the strong acceleration followed by an even stronger deceleration is caused by the dynamics of the phase-field.
This process of deceleration for larger amounts of liquids, \(c=0.2\) and \(c=0.3\), can be explained as the effect of drag (see Sec.~\ref{sec:results_dynamics_two}).
In Fig.~\ref{fig:spheres}, one can see that the effect of drag is more significant for larger amounts of liquid, as to be expected.

By looking at the distance or velocity profiles in Fig.~\ref{fig:spheres}, one can easily identify the time when the particles are in contact.
In the velocity profile, this point in time is indicated by the steep drop of the velocity to zero.
The contact time is the shortest for a liquid fraction of \(c=0.3\) and the largest for \(c=0.5\).
This observation is also true for the three and four-particle systems.

In the second setup, we placed three spherical particles, with the same radius and distance as in the first case, in such a way that they form an equilateral triangle.
Different amounts of liquid are initialized in the center of the triangle so that the resulting liquid volume fractions are the same as in the first case.
Fig.~\ref{fig:spheres} shows the decreasing average distance \(\tilde h = 1/3 \sum_i h_i\) between the three particles.
Because the particles form an equilateral triangle, the force on each particle is the same, and hence the particles form an equilateral triangle during the entire simulation.
This means that the average distance is equal to the distance between the particle \(\tilde h = h_2 = h_3 = h_3\).

The results we obtain for three particles are qualitatively similar to the results obtained for two particles.
However, one can see that the time, until the particles come into contact, is reduced for \(c=0.2\) and \(c=0.3\).
In the case of \(c=0.5\), the effect of drag is stronger than in the case of two particles.

In the last setup, we placed four spherical particles, with the same radius and distance as in the first case, in such a way that they form a tetrahedron.
Again, we have initialized spherical liquid drops of different sizes in the center of the tetrahedron and waited for ten thousand simulation time steps so that the liquid reaches its equilibrium before the particles are allowed to move (\(t=0\)).
Likewise, Fig.~\ref{fig:spheres} shows the decreasing average distance \(\tilde h = 1/4 \sum_i h_i\) between the particles for different liquid volume fractions.

The results we obtain for four particles are qualitatively similar to the results obtained for two and three particles.
The time until the particles come into contact is reduced for \(c=0.2\) and \(c=0.3\) compared to two particles but is comparable to the case of three particles.
Furthermore, the above-discussed effect of drag is even stronger than in the case of three particles.
The fact that the effect of drag increases with the number of particles can be explained by the higher amount of solid surface in contact with the liquid which leads to a higher drag force.

\section{Summary and Conclusion}
\label{sec:summary}

We present a combined phase-field-lattice Boltzmann model for the simulation of liquid state sintering.
The accuracy of the present method is carefully investigated by considering a liquid bridge between two parallel plates of finite size.
We show that the obtained results for their motion are in agreement with the present theoretical predictions.
The capability of the present model to simulate liquid state sintering is demonstrated by the investigation of the dynamics of two, three, and four solid particles under the action of a capillary attraction between them.
In all the cases investigated, increasing the amount of liquid first accelerates the compaction process.
A liquid fraction higher than a certain threshold, however, slows down the motion of spheres and leads to an increase of the time necessary for the particles to come into contact.
We show that this is caused by the dynamics of the fluid, i.e.\ viscous drag.
There is thus an optimum choice for liquid content concerning the compaction process.
The exact value of this optimum parameter will depend in general on the powder packing fraction, grain shape, and size distribution.
The present method provides a versatile tool to explore this important issue.

\section*{Acknowledgments}
Financial support by the German Research Foundation DFG under the grant VA205/17-1 is gratefully acknowledged.

\appendix

\section{Liquid Bridge Between two Plates --- Force}
\label{app:theory_plate_force}

We consider a cylindrical capillary between two finite plates.
As shown above, the change of internal energy is given by
\begin{math}
    \mathrm d U =  \sigma \,\mathrm d A
\end{math}
with \(A=2\pi R\) one obtains
\begin{math}
    \mathrm d U =  2\pi\sigma \,\mathrm d R
\end{math}
Further, we consider the energy of the liquid phase \(U_{{L}}\) as a function of the distance \(h_c\) between the two plates.
Because the liquid volume is constant \(V_L = \pi R_0^2 h_{0} = \pi R^2 h_c\), the radius \(R\) can be expressed as function of the initial radius \(R_0\), initial distance \(h_{0}\) and distance \(h_c\) with \(R=R_0\sqrt{h_{0}/h_c}\).
Hence, the force is obtained as
\begin{dmath}
    {F= \left(\frac{\mathrm d U}{\mathrm d h_c}\right)_{V_L} = \pi\sigma R_0\sqrt{h_{0}/h_c}}
\end{dmath}.

\section{Liquid Bridge Between two Solid Spheres}
\label{app:theory_2_spheres}

\subsection{Spherical Bridge}
Equation~(\ref{eq:theory_2_spheres_spherical}) describes the necessary wetting angle to form spherical capillary bridge as a function of the liquid fraction.
To derive this relation, we consider two spherical solid bodies in contact, as depicted in Fig.~\ref{fig:bridge_spherical}.

\subsubsection{Small Spherical Capillary Bridge}
\label{app:small_spherical}

\begin{figure}
    \includegraphics{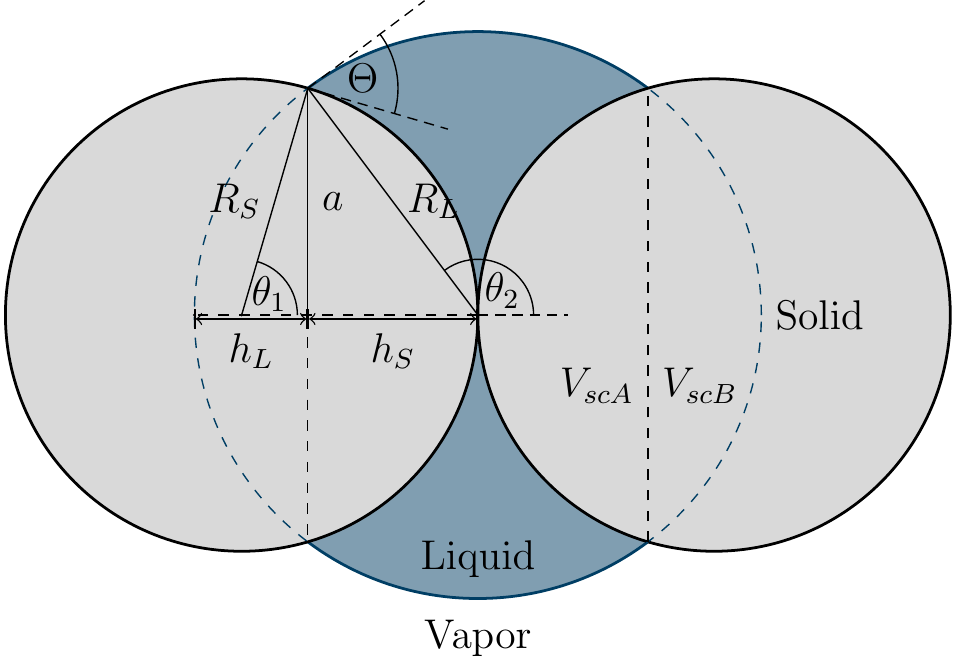}
    \caption{Depicted are two spherical solid bodies with equal radii \(R_{{S}}\).
    The two bodies are in contact with each other and are connected by a liquid spherical capillary bridge with the radius \(R_{{L}} = \chi_{{S}} R_{{S}}\).
    The figure has been generated with the parameter \(\chi_S = 1.2\).
    }
    \label{fig:bridge_spherical}
\end{figure}

The contact angle \(\Theta_{{S}}\) is defined as the angle between the tangent lines of the liquid and solid surface at the triple point.
It is thereby equal to the angle between the surface normals vectors of the liquid and solid surface at the triple point and can be calculated with

\begin{dmath}
    \Theta_{{S}} = \theta_2 - \theta_1
\end{dmath},
where \(\theta_1\) and \(\theta_2\) are the angles between the surface normals and the depicted dashed line (see. Fig.~\ref{fig:bridge_spherical}).
These angles can be calculated by

\begin{dgroup}
    \begin{dmath}
       \theta_1 = \arccos\left(\frac{R_{{S}} - h_{{S}}}{R_{{S}}}\right)
    \end{dmath}
    \begin{dmath}
       \theta_2 = \pi - \arccos\left(\frac{h_{{S}}}{\chi_{{S}} R_{{S}}}\right)
    \end{dmath},
    \label{eq:app_bs_angles}
\end{dgroup}
where \(\chi_{{S}} \in \left[0,2\right]\) is a dimensionless scaling factor.
We introduced this scaling factor in order to relate the radius of the capillary bridge \(R_{{L}}\) to the radius of the solid spheres \(R_{{S}}\) with \(R_{{L}} = \chi_{{S}} R_{{S}}\).
Again, one can obtain a system of equations for the side lengths of the depicted triangles by using the Pythagorean theorem,

\begin{dgroup}
    \begin{dmath}
        \chi_{{S}} R_{{S}} = h_{{L}} + h_{{S}}
    \end{dmath}
    \begin{dmath}
        R_{{S}}^2 = a^2 + \left(R_{{S}} - h_{{S}}\right)^2
    \end{dmath}
    \begin{dmath}
        \chi_{{S}}^2 R_{{S}}^2 = a^2 + h_{{S}}^2
    \end{dmath},
    \label{eq:app_bs_soe_1}
\end{dgroup}
with the solution

\begin{dgroup}
    \begin{dmath}
        h_{{S}} = \frac{\chi_{{S}}^2}{2} R_{{S}}
        \label{eq:b_A}
    \end{dmath}
    \begin{dmath}
        h_{{L}} = \frac{2\chi_{{S}}-\chi_{{S}}^2}{2} R_{{S}}
        \label{eq:b_B}
    \end{dmath}
    \begin{dmath}
        a = \frac{1}{2}\sqrt{4\chi_{{S}}^2-\chi_{{S}}^4} R_{{S}}
    \end{dmath}.
    \label{eq:app_parameters}
\end{dgroup}
In Eq.~(\ref{eq:app_parameters}), we expressed the geometric parameters depicted in Fig.~\ref{fig:bridge_spherical} as function of the radius \(R_{{S}}\) and the scaling parameter \(\chi_{{S}}\).
As to be expected, the contact angle can be expressed as a function of \(\chi_{{S}}\) with

\begin{dmath}
    \Theta_{{S}}
    = \arccos\left(\frac{\chi_{{S}}^2}{2} - 1\right) - \arccos\left(\frac{\chi_{{S}}}{2}\right)
    \label{eq:app_theta_chi_spherical}
\end{dmath}.

Furthermore, we replace the scaling factor \(\chi_{{S}}\) by the liquid volume fraction \(c\).
Therefore, the volume of the liquid has to be calculated by subtracting the volume of the spherical caps \(V_{{scA}}\) and \(V_{{scB}}\) (see Fig.~\ref{fig:bridge_spherical}) form the volume of a liquid sphere \(V_{{Ls}}\) with

\begin{dmath}
    V_{{L}} = V_{{Ls}}  - 2 V_{{scA}} - 2 V_{{scB}}
    \label{eq:v_l}
\end{dmath}
and

\begin{dgroup}
    \begin{dmath}
        V_{{Ls}} = \frac{4\pi}{3} \chi_{{S}}^3 R^3
        \label{eq:v_ls}
    \end{dmath}
    \begin{dmath}
        V_{{scA}} = \frac{\pi}{3} h_{{S}}^2 \left(3R_{{S}} - h_{{S}}\right)
        \label{eq:v_scA}
    \end{dmath}
    \begin{dmath}
        V_{{scB}} = \frac{\pi}{3} h_{{L}}^2 \left(3\chi_{{S}} R_{{S}}-h_{{L}}\right)
        \label{eq:v_scB}
    \end{dmath}
\end{dgroup}
and Eq.~(\ref{eq:app_parameters}).
The volumes of the spherical caps, \(V_{{scA}}\) and \(V_{{scB}}\), can be parametrized with \(R_{{S}}\) and \(\chi_{{S}}\) by inserting Eq.~(\ref{eq:b_A}) into Eq.~(\ref{eq:v_scA}) and Eq.~(\ref{eq:b_B}) into Eq.~(\ref{eq:v_scB}) so that

\begin{dgroup}
    \begin{dmath}
        V_{{scA}} = \pi R_{{S}}^3 \left(-\frac{1}{24}\chi_{{S}}^6 + \frac{1}{4} \chi_{{S}}^4 \right)
        \label{eq:v_scA_chi}
    \end{dmath}
    \begin{dmath}
        V_{{scB}}
        =
        \pi R_{{S}}^3 \left(\frac{1}{24} \chi_{{S}}^6 - \frac{1}{2} \chi_{{S}}^4 + \frac{2}{3}\chi_{{S}}^3\right)
        \label{eq:v_scB_chi}
    \end{dmath}.
\end{dgroup}
Inserting Eqs.~(\ref{eq:v_scA_chi}), (\ref{eq:v_scB_chi}), and (\ref{eq:v_ls}) into Eqs.~(\ref{eq:v_l}) delivers a simple parametrization for the liquid volume

\begin{dmath}
    V_{{L}} = \frac{1}{2} \pi \chi_{{S}}^4 R_{{S}}^3
\end{dmath}.
The same way as for the cylindrical bridge, we can calculate the liquid volume fraction with

\begin{dgroup}
    \begin{dmath}
        c = \frac{V_{{L}}}{V_{{L}} + V_{{S}}}
    \end{dmath}
    \begin{dmath}
          = \frac{3\chi_{{S}}^4}{3\chi_{{S}}^4 + 16}
    \end{dmath}.
    \label{eq:app_c_chi}
\end{dgroup}
By rearranging the terms of Eq.~(\ref{eq:app_c_chi}), we obtain a fourth order polynomial equation for \(\chi_{{S}}\)

\begin{dmath}
    \chi_{{S}}^4 = \frac{16c}{3\left(1-c\right)}
\end{dmath},
which has a real and positive solution

\begin{dmath}
    \chi_{{S}}
    = \frac{2}{\sqrt[4]{3}}\frac{\sqrt{\left(1-c\right)\sqrt{c\left(1-c\right)}}}{1-c}
    \label{eq:app_chi_s}
\end{dmath}.
Inserting Eq.~(\ref{eq:app_chi_s}) into Eq.~(\ref{eq:app_theta_chi_spherical}) delivers the solution shown in Fig.~\ref{fig:theory_2spheres_phases}.

\subsubsection{Large Spherical Capillary Bridge}
\begin{figure}
    \includegraphics{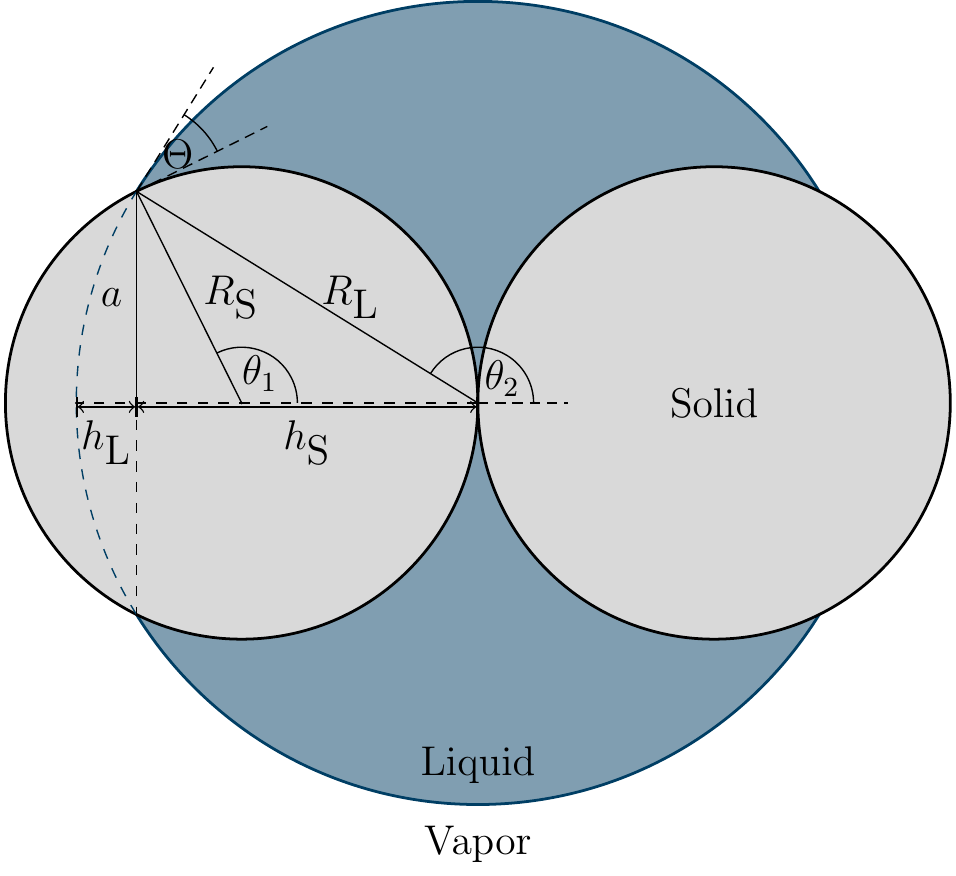}
    \caption{Depicted are two spherical solid bodies with equal radii \(R_{{S}}\).
    The bodies are in contact with each other and are connected by a liquid spherical capillary bridge with the radius \(\chi_{S} R_{{S}}\) and the contact angle \(\Theta\).
    The figure has been generated with the parameter \(\chi_S = 1.7\).
    }
    \label{fig:bridge_spherical_large}
\end{figure}

In the case of relatively large amount of liquid, a liquid bridge as depicted in Fig.~\ref{fig:bridge_spherical_large} may form.
The angles \(\theta_1\) and \(\theta_2\) are obtained in the same manner as above with

\begin{dgroup}
    \begin{dmath}
       \Theta_{{S}} = \theta_2 - \theta_1
    \end{dmath}
    \begin{dmath}
       \theta_1 = \pi - \arccos\left(\frac{h_{{S}} - R_{{S}}}{\chi_{{S}} R_{{S}}}\right)
    \end{dmath}
    \begin{dmath}
       \theta_2 = \pi - \arccos\left(\frac{h_{{S}}}{R_{{S}}}\right)
    \end{dmath},
    \label{eq:app_bs_theta}
\end{dgroup}
where \(\theta_1\) can be simplified to

\begin{dmath}
    \theta_1 = \arccos\left(\frac{R_{{S}} - h_{{S}}}{\chi_{{S}} R_{{S}}}\right)
\end{dmath}.
The height \(h_{{S}}\) can be calculated in the same manner as in~\ref{app:small_spherical}, and Eq.~(\ref{eq:app_theta_chi_spherical}) can be obtained again.

\subsection{Cylindrical bridge}

\begin{figure}
    \includegraphics{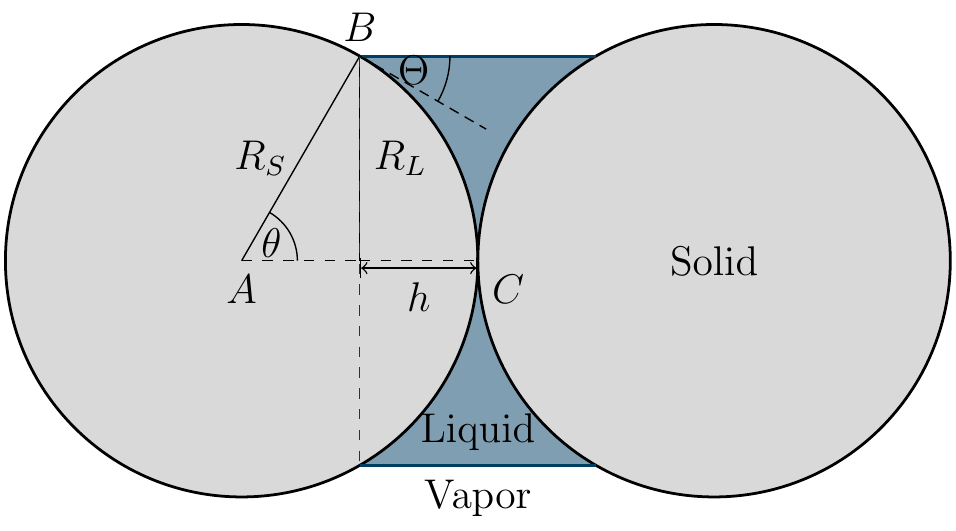}
    \caption{Depicted are two spherical solid bodies with the radius \(R_{{S}}\).
    The two bodies are in contact with each other and are connected by a liquid cylindrical capillary bridge with the radius \(R_{{L}}\), the height \(2h\), and the contact angle \(\Theta\).}
    \label{fig:bridge_cylinderXY}
\end{figure}

Equation~(\ref{eq:theory_2_spheres_cylindrical}) describes the wetting angle necessary to form a cylindrical capillary bridge as a function of the liquid fraction.
To derive this relation, we consider two spherical solid bodies in contact, as depicted in Fig.~\ref{fig:bridge_cylinderXY}.

To calculate the contact angle \(\Theta_{{c}}\), we consider the angle \(\theta\) between two lines.
The first line starts at the center of the sphere \(A\) ends in the triple point \(B\).
The second line also starts at the center of the sphere \(A\) but ends in the contact point \(C\) between the spheres so that

\begin{math}
    \theta = \arccos{\left(\frac{R-h}{R}\right)}
\end{math},
where \(h\) is half of the height of the cylindrical capillary bridge.
By considering the intercept theorem, we obtain the contact angle

\begin{dmath}
    \Theta_{{c}} = \frac{\pi}{2} - \arccos{\left(1-\chi_{{c}}\right)}
    \label{eq:app_theta_cylindrical}
\end{dmath},
where we have introduced the dimensionless scaling parameter \(\chi_{{c}} \in \left[0,1\right]\) with \(h=\chi_{{c}} R\).

The volume of the liquid can be determined by calculating the volume of a
cylinder and subtraction of the solid sphere caps, which cover part of the cylinder's volume,

\begin{dgroup}
\begin{dmath}
    V_{{L}}
    = V_{{cylinder}} - 2 V_{{sphere-cap}}
\end{dmath}
\begin{dmath}
    = 2 \pi R_{{S}}^2 h -
      2 \frac{\pi}{6} h \left[ 3R_{{S}}^2 + h^2 \right]
\end{dmath}
\begin{dmath}
    = \pi R_{{S}}^2 h - \frac{\pi}{3} h^3
    \label{eq:volume_liquid}
\end{dmath}.
\end{dgroup}
By using the Pythagorean theorem
\begin{math}
    R^2 = R_{{S}}^2 + \left(R-h\right)^2
\end{math},
the radius of the capillary bridge \(R_{{S}}\) in Eq.~(\ref{eq:volume_liquid}) can be replaced by the radius of the solid sphere \(R_{{S}}\).
With this replacement, we obtain for the fluid volume

\begin{dmath}
    V_{{L}}
    = 2 \pi R h^2 - \frac{4}{3} \pi h^3
\end{dmath}.
Furthermore, we replace height \(h\) again with \(h= \chi_{{c}}R\) so that the expression for the fluid volume

\begin{dmath}
    V_{{L}}
    = \left(- \frac{4}{3} \chi_{{c}}^3 + 2 \chi_{{c}}^2\right) \pi R^3
    \label{eq:volume_liquid_2}
\end{dmath}
can be further simplified.
With Eq.~(\ref{eq:volume_liquid_2}) and the volume of the solid,
\begin{math}
    V_{{S}} = \frac{8}{3} \pi R^3
\end{math},
the liquid volume fraction can be calculated

\begin{dgroup}
\begin{dmath}
    c
    = \frac{V_{{L}}}{V_{{L}} + V_{{S}}}
\end{dmath}
\begin{dmath}
    = \frac{2\chi_{{c}}^3 - 3 \chi_{{c}}^2}
      {2\chi_{{c}}^3 - 3 \chi_{{c}}^2 - 4}
    \label{eq:cylinder_liquid_fraction}
\end{dmath}.
\end{dgroup}
Equation~(\ref{eq:cylinder_liquid_fraction}) can be transformed in a third order polynomial

\begin{dmath}
    \chi_{{c}}^3 - \frac{3}{2}\chi_{{c}}^2 + \frac{2c}{1-c}= 0
    \label{eq:tilde_h_poly}
\end{dmath}.
The polynomial equation~(\ref{eq:tilde_h_poly}) has three real roots for \(c\in\left(0,0.2\right)\).
By substituting \(\chi_{{c}} = t + \frac{1}{2}\), we can write Eq.~(\ref{eq:tilde_h_poly}) in the form,

\begin{dmath}
    t^3 + p t + q =0
    \label{eq:ploynom_pq}
\end{dmath}
with
\begin{math}
    p = -\frac{1}{4}
\end{math}
and
\begin{math}
    q = \frac{2c}{1-c} -\frac{1}{4}
\end{math}.
With the use of the cosine power rule

\begin{dmath}
    \cos^3\left(\eta\right)
    - \frac{3}{4} \cos\left(\eta\right)
    - \frac{1}{4} \cos\left(3\eta\right)
    = 0
\end{dmath},
we find three real solutions of Eq.~(\ref{eq:ploynom_pq})

\begin{dmath}
    t_n
    =
    \sqrt{-\frac{4p}{3}} \cos
    \left(
        \frac{1}{3}\arccos
        \left[
            -4q\left(-\frac{4p}{3}\right)^{\frac{3}{2}}
        \right]
        + \frac{2n\pi}{3}
    \right)
\end{dmath}
with \( n \in \left\{0,1,2\right\}\).
By back substitution, we also obtain three real solutions for the scaling factor

\begin{dmath}
    \chi_{{c}n}
    =
    -\cos
    \left(
        \frac{1}{3}\arccos
        \left(
            \frac{1-9c}{1-c}
        \right)
        + \frac{2n\pi}{3}
    \right)
    + \frac{1}{2}
    \label{eq:app_chi_c}
\end{dmath}.
The physically meaningful solution of \(\chi_{{c}}\) has to be limited to the interval \(\left[0,1\right]\), which is only fulfilled by \(\chi_{{c}2}\).
Inserting Eq.~(\ref{eq:app_chi_c}) into Eq.~(\ref{eq:app_theta_cylindrical}) delivers the result shown in Fig.~\ref{fig:theory_2spheres_phases}.

\bibliography{Zotero.bib}

\end{document}